\documentclass[useAMS,usenatbib]{mn2e}

\usepackage{txfonts}
\usepackage{longtable}

\usepackage{graphicx}
\usepackage{longtable}
\usepackage{appendix}
\usepackage{lscape}
\newcommand{\mlu}{\mbox{$M_{\odot}$\,yr$^{-1}$}}

\title[LMC post-AGB stars]{
{\it Spitzer Space Telescope} spectra of post-AGB stars in the Large Magellanic Cloud 
---polycyclic aromatic hydrocarbons
at low metallicities  }
   
\author[Matsuura et al.]
{Mikako Matsuura$^{1}$,
Jeronimo Bernard-Salas$^{2, 3}$,
T. Lloyd Evans$^{4}$, 
Kevin M. Volk$^{5}$,   \newauthor  
Bruce J. Hrivnak$^{6}$, 
G.C. Sloan$^{7}$,
You-Hua Chu$^{8}$,
Robert Gruendl$^{8}$,
Kathleen E. Kraemer$^{9}$,    \newauthor 
Els Peeters$^{10,11}$,       
R. Szczerba$^{12}$,
P.R. Wood$^{13}$,
Albert A. Zijlstra$^{14}$,  
S. Hony$^{15}$,   
Yoshifusa Ita$^{16}$,  \newauthor 
Devika Kamath$^{13}$, 
Eric Lagadec$^{7}$, 
Quentin A Parker$^{17, 18, 19}$,   
Warren A. Reid$^{17}$,    \newauthor 
Takashi Shimonishi$^{20}$,   
H. Van Winckel$^{21}$,    
Paul M. Woods$^{1,22}$, 
F. Kemper$^{23}$, 
Margaret Meixner$^{5}$,\newauthor
M. Otsuka$^{23}$, 
R. Sahai$^{24}$, 
B.A. Sargent$^{25}$,  
J.L. Hora$^{26}$, 
Iain McDonald$^{14}$, \\
$^{1}$ Department of Physics and Astronomy, University College London, Gower Street, London WC1E 6BT, UK\\
$^{2}$ Department of Physical Sciences, The Open University, Milton Keynes, MK7 6AA, UK \\
$^{3}$ Institut d'Astrophysique Spatiale, CNRS/Universit{\'e} Paris-Sud 11, 91405 Orsay, France \\
$^{4}$ SUPA, School of Physics and Astronomy, University of St Andrews, North Haugh, St Andrews, Fife, Scotland KY16 9SS, UK \\
$^{5}$ Space Telescope Science Institute, 3700 San Martin Drive, Baltimore, MD 21218, USA \\
$^{6}$ Department of Physics and Astronomy, Valparaiso University, Valparaiso, IN 46383, USA \\
$^{7}$ Department of Astronomy, Cornell University, Ithaca, NY 14853, USA \\
$^{8}$ Department of Astronomy, University of Illinois at Urbana-Champaign, 1002 West Green Street, Urbana, IL 61801, USA \\
$^{9}$ Institute for Scientific Research, Boston College, Kenny Cottle L106B, Newton, MA 02459-1161, USA \\
$^{10}$ Department of Physics and Astronomy, University of Western Ontario, London, Ontario N6A 3K7, Canada \\
$^{11}$ SETI Institute, 515 North Whisman Road, Mountain View, CA 94043, USA \\
$^{12}$ N. Copernicus Astronomical Center, Rabianska 8, 87-100 Torun, Poland \\
$^{13}$Research School of Astronomy and Astrophysics, Australian National University, Weston Creek, ACT 2611, Australia\\
$^{14}$ Jodrell Bank Centre for Astrophysics, School of Physics and Astronomy, University of Manchester, Oxford Road, Manchester M13 9PL, UK\\
$^{15}$ Max-Planck-Institute for Astronomy, K\"{o}nigstuhl 17, 69117 Heidelberg, Germany \\
$^{16}$ Astronomical Institute, Graduate School of Science, Tohoku University, 6-3 Aramaki Aoba, Aoba-ku, Sendai, Miyagi 980-8578, Japan \\
$^{17}$ Department of Physics \& Astronomy, Macquarie University, NSW 2109, Australia \\
$^{18}$ Research Centre in Astronomy, Astrophysics and Astrophotonics (MQAAAstro), Macquarie University, NSW 2109, Australia \\
$^{19}$ Australian Astronomical Observatory, PO Box 296, Epping, NSW 2121, Australia \\
$^{20}$ Department of Earth and Planetary Sciences, Graduate School of Science, Kobe University, Nada Kobe 657-8501, Japan \\
$^{21}$ Instituut voor Sterrenkunde, K.U.Leuven, Celestijnenlaan 200B, B-3001 Leuven, Belgium \\
$^{22}$ Astrophysics Research Centre, School of Mathematics and Physics, Queen's University Belfast, Belfast BT7 1NN, UK \\
$^{23}$ Academia Sinica, Institute of Astronomy and Astrophysics, 11F of ASMAB, AS/NTU, 
              No. 1, Sec. 4, Roosevelt Rd, Taipei 10617, Taiwan, R.O.C. \\
$^{24}$ Jet Propulsion Laboratory, MS 183-900, California Institute of Technology, Pasadena, CA 91109, USA \\
$^{25}$ Center for Imaging Science, Rochester Institute of Technology, 54 Lomb Memorial Drive, Rochester, NY 14623, USA \\
$^{26}$ Harvard-Smithsonian Center for Astrophysics, 60 Garden Street, MS 65, Cambridge, MA 02138-1516, USA \\
 }
\date{Released 2013; to be published in 2014}

\pagerange{\pageref{firstpage}--\pageref{lastpage}} \pubyear{2013}

\def\LaTeX{L\kern-.36em\raise.3ex\hbox{a}\kern-.15em
    T\kern-.1667em\lower.7ex\hbox{E}\kern-.125emX}

\usepackage{mn2e-breakabs}
\begin{document}

\label{firstpage}

\maketitle

\begin{abstract}

This paper reports variations of polycyclic aromatic hydrocarbons (PAHs) features that were found in {\it Spitzer Space Telescope} spectra of carbon-rich post-asymptotic giant branch (post-AGB) stars 
in the Large Magellanic Cloud (LMC).  
The paper consists of two parts. 
The first part describes our {\it Spitzer} spectral observing programme of 24 stars including post-AGB candidates. 
%
The latter half of this paper presents the analysis of PAH features in 20 carbon-rich post-AGB stars in the LMC, assembled from the Spitzer archive as well as from our 
own programme.
We found that five post-AGB stars showed a broad feature with a peak at 7.7\,$\mu$m, that had not been classified before.
Further, the 10--13\,$\mu$m PAH spectra were classified into four classes, one of which has three broad peaks at 11.3, 12.3 and 13.3\,$\mu$m rather than two 
distinct sharp peaks at 11.3 and 12.7\,$\mu$m, as commonly found in HII regions.
Our studies suggest that PAHs are gradually processed while the central stars evolve from post-AGB phase to PNe, changing their composition before PAHs are incorporated 
into the interstellar medium. Although some metallicity dependence of PAH spectra exists,
 the evolutionary state of an object is more significant than its metallicity in determining the spectral characteristics of PAHs for LMC and Galactic post-AGB stars.
\end{abstract}



\section{Introduction}

The Large Magellanic Cloud (LMC) has attracted much attention in recent years, as its relative proximity \citep[50\,kpc; ][]{Westerlund97} and relatively low metallicity 
\citep[about half of the solar metallicity; ][]{Olszewski:1996ji} make the LMC an important laboratory to examine the effects of metallicity on molecular and dust 
formation.

It is increasingly important to understand how different metallicities could impact on the physics and chemistry of the constituent stars and the interstellar medium 
(ISM) of galaxies. 
Different metallicities might result in different compositions of dust and molecules, changing the spectra of stars and the ISM as well as those of the integrated 
light of galaxies. 
A set of features which are often found in the spectra of stars, the ISM and galaxies have been attributed to polycyclic aromatic hydrocarbons (PAHs). 
\citet{Madden:2006fj} found that PAH features were less pronounced in low-metallicity dwarf galaxies.
Similarly, the Spitzer SINGS survey of 67 galaxies within 3--25 Mpc \citep{KennicuttJr:2003jt} showed that PAH features were weaker in galaxies of lower metallicity 
\citep{Draine:2007de}.
In contrast, PAH features of planetary nebulae (PNe) showed no difference between the Milky Way, the LMC and the Small Magellanic Cloud  (SMC)
\citep{BernardSalas:2009iq}, despite the LMC and the SMC having about half and a quarter of the solar metallicity, respectively.
It is still unclear how low metallicities affect PAHs in stars and the ISM in a different manner.

It has been proposed that carbon-rich asymptotic giant branch (AGB) stars are an ideal site for PAH formation \citep{1989ApJS...71..733A}.
Carbon atoms are synthesised in stellar interiors during the AGB phase, and are dredged-up to the outer layers of the star \citep{Herwig:2005jn}.
Although the atmospheres of AGB stars contain more oxygen than carbon initially, successive dredge-ups will increase the abundance of carbon, resulting in more carbon 
than oxygen in some stars within a certain mass range \citep{Vassiliadis:1993jk}.
Carbon and oxygen atoms in stellar atmospheres bond to form CO first, and excess carbon atoms are available to form other carbon-bearing molecules.
One of the key molecules in carbon-rich chemistry is C$_2$H$_2$, which is thought to be the parent molecule in the formation of PAHs \citep{1989ApJS...71..733A, 
1992ApJ...401..269C}. A sequence of C$_2$H$_2$ added onto carbon and a hydrogen chain forms aromatic hydrocarbons; further additions of C$_2$H$_2$ and 
C$_4$H$_2$ form larger PAHs. As the abundance of C$_2$H$_2$ in AGB atmospheres is metallicity dependent \citep{Matsuura:2005eja}, the formation of PAHs in 
carbon-rich AGB stars could also depend on metallicity.

In general, PAHs are rarely seen in the spectra of AGB stars \citep{Sloan:2007hf}.
Only when the effective temperature of the central star increases in the post-AGB phase and subsequent PN phase do PAHs appear in their spectra. 
This is because PAHs require energy supplied by UV or optical radiation to emit their infrared spectra \citep{Leger:1984vj, 1989ApJS...71..733A, Sloan:2007hf}.
Thus, in order to investigate PAHs immediately after their formation, we have chosen to study PAHs in post-AGB stars.

PAHs exhibit a variety of spectral features.
The {\it Infrared Space Observatory} (ISO) observations of Galactic objects showed that PAH spectral profiles in the 6--9 and 10--15\,$\mu$m regions vary among objects 
\citep[e.g. HII regions, PNe, post-AGB stars; ][]{Peeters:2002ci, Hony:2001iw}.
In the 10--15\,$\mu$m region, there are two strong peaks located at 11.3\,$\mu$m and 12.7\,$\mu$m, and the relative intensity of the 11.3\,$\mu$m feature over that at 
12.7\,$\mu$m spans a factor of 7. 
\citet{Hony:2001iw} suggested that the change of the 10--15\,$\mu$m profile should be caused by the different structures and compositions of PAHs. 
In the meantime, \citet{Peeters:2002ci} found most profiles of the 6--9\,$\mu$m features fall into three groups, which are correlated with object type.
\citet{Peeters:2002ci} suggested that nitrogenation causes the shift of the peaks of the 6--9\,$\mu$m feature.
Alternatively, it has been proposed that aliphatic carbon could further contribute to the spectral variations in the 6--9\,$\mu$m feature \citep[e.g.][]
{Sloan:2007hf, Kwok:2011iv, Carpentier:2012cr}.
Further, hydrogenation should also contribute to the variation \citep{Duley:1981we, Jones:1990vd}. 
The cause of spectral variations at 6--9\,$\mu$m is still unsettled.

 This paper focuses on mid-infrared spectra of PAHs found in post-AGB stars in the LMC. 
Our objectives are  to search for any difference in PAH profiles between the LMC and the Milky Way (MW), and to examine whether the differences in PAH profiles may be 
caused by the metallicity difference. 
The first part (Sect. 2--4) of this paper outlines our {\it Spitzer} observing programme of post-AGB stars in the LMC. 
We selected post-AGB candidates in the LMC using several different criteria, including position in colour-colour and colour-magnitude diagrams, and we will verify these 
selection methods. 
The second part (Sect. 5) is the investigation of PAH features in post-AGB stars. 
For this analysis, archival data are added in addition to our own {\it Spitzer} observations. 
LMC post-AGB stars show more spectral variations of PAHs than are known from Galactic counterparts.
We discuss the possible causes of these variations.

\section{Spitzer cycle-5 observing programme}

 \subsection{Target selection} \label{targets}
 
   Our primary targets were post-AGB candidates in the LMC. 
Additionally we included a few candidate AGB stars and PNe, in order to examine the spectral evolution from the AGB phase through the post-AGB phase to the PN phase.

 Although theoretically the post-AGB phase is well defined as the transition phase between the AGB and the PN phases, there are several observational characteristics 
that could match the theoretical definition.
We employed three different selection approaches for post-AGB candidates:
   a) a selection based on infrared colour-colour magnitude diagrams; 
   b) a selection of stars with double-peaked spectral energy distributions (SEDs);
   c) a selection of stars which were identified as post-AGB candidates in the literature. 
The identifications of post-AGB stars in the literature were based on combinations of infrared colours, optical spectra, luminosities and  light curves.
Within our team, we searched for candidates independently with five different selection criteria, and created a combined sample.
The selection criteria are briefly described below, and the method applied to a specific target is indicated in Table\,\ref{table-targets}.
Later in this paper, we will re-examine these selection criteria based on colours. 
We restricted our sample to [8.0]$<$8, where [8.0] is the magnitude at 8.0\,$\mu$m, in order to obtain a good signal to noise ratio.

\begin{enumerate}

  \item 
  Extremely Red Objects (EROs):   
  As their names imply, EROs are very red at mid-infrared wavelengths. \citet{Gruendl:2008fy} defined EROs to have { \it Spitzer} colours of [4.5]$-$[8.0]$>4.0$, 
where [4.5] and [8.0] are magnitudes in the 4.5\,$\mu$m and 8.0\,$\mu$m bands in the {\it Spitzer} LMC Point Source Catalog \citep[SAGE; ][]{Meixner:2006eg}.
 These EROs are stars which are heavily obscured by dust, and they are probably  related with mass loss at the late AGB phase.
  Some of the EROs show double-peaked SEDs \citep{Gruendl:2008fy}, suggesting that the extensive mass-loss rate of the AGB phase has terminated, and the star has entered 
the post-AGB phase.  Five EROs were observed.

  \item 
  \citet{2011A&A...530A..90V} selected 70 probable post-AGB candidates.
   They initially selected stars as potential candidates on the basis of their $[8]-[24]$ colour, and narrowed down the list of candidates
   by the SEDs, luminosities and optical spectra.
   From these candidates, we chose one source with a `circumbinary disk' (J053336.37$-$692312.7), and one source with a `steady state expanding shell' 
(J052043.58$-$692341.4).
   
   \item 
Galactic post-AGB stars can indicate the typical infrared colours expected for LMC post-AGB stars.
\citet{Szczerba:2007eb} assembled a list of Galactic post-AGB stars, some of which have ISO/SWS spectra \citep{Sloan:2003ki}.
These spectra were convolved with the {\it Spitzer} and 2MASS filter transmission curves, producing synthetic infrared colours of post-AGB objects.
We plotted K$-$[24] and $K-[8.0]$ colours of Galactic post-AGB stars and 
selected  two SAGE LMC post-AGB candidates which have similar colours to Galactic post-AGB stars. 
    
\item 
Two stars (MSX\,SMC\,29 and SMP\,LMC\,11) in the Magellanic Clouds were identified as carbon-rich post-AGB stars based on their Spitzer spectra
\citep{Kraemer:2006cl, BernardSalas:2006ck}.
Their magnitudes and colours indicated  [8]$<$8, $J-[8.0]>5$ and $[8]-[24]>0.5$, and we selected  seven stars which met these criteria from the SAGE catalogue 
\citep{Meixner:2006eg}.
   
 \item 
  \citet{2001ASSL..265...71W} identified LMC post-AGB candidates from the MSX infrared survey  \citep{Egan:2001tl}
  and MACHO light curves \citep{Alcock:2001fq}. 
We included five targets from \citet{2001ASSL..265...71W},
  These targets were selected for a variety of reasons.  MSX LMC 130 had an intermediate oxygen-rich optical spectral with emission lines.
MSX LMC 736 showed an early M spectral-type, slowly declining magnitudes and a very bright mid-IR SED 
though later the M-type star turned out to be a foreground M dwarf coincident with a dusty carbon-rich star.
MSX LMC 439 is an R CrB star with a characteristic
R CrB light curve, and the other two stars had near- to mid-IR SEDs
indicating likely binary post-AGB stars \citep{VanWinckel:2003iw}.

\end{enumerate}

In addition to the above selections for low-resolution spectral targets, we selected three targets for high-resolution spectra, in order to identify narrow molecular and 
ice features.
\citet{Kemper:2010bwa} have taken low-resolution IRS spectra of about 200 point sources in the LMC.
Low-resolution spectra of one object (IRAS 04518$-$6852) gave a hint of molecular features from 10--15 \,$\mu$m, so we requested follow-up observations with higher 
spectral resolution.
The second is a young stellar object (YSO),  J\,053941$-$692916 identified by an ice band in an AKARI spectrum \citep{Shimonishi:2010ek}.
The last one (SMP\,LMC\,75) had a similar mid-infrared colour to a molecular-rich post-AGB star in the LMC \citep{Kraemer:2006cl}.


 \subsection{Observations and data reduction}

We observed 24 post-AGB candidates, late-phase AGB stars and PNe with the Infrared Spectrograph \citep[IRS; ][]{Houck:2004el} on board the {\it Spitzer Space Telescope}  
\citep{Werner:2004jt}.
The programme ID was 50338. The observations were made between 2009 March 6th and 2009 April 29th.
We mainly used the low-resolution IRS modules, consisting of Short-Low (SL) and Long-Low (LL), except for three sources, which were observed with the 
high-resolution modules using Short-High (SH) and Long-High (LH).
Table\,\ref{table-targets} gives the dates and integration times for each module,
SAGE coordinates and AOR (Astronomical Observation Request) numbers.

The data were processed using the {\it Spitzer} Science Center's (SSC's) pipeline (version 18.7), which was maintained at Cornell, with the Smart reduction package 
\citep{Higdon:2004iy, Lebouteiller:2010dw}.
The basic calibrated data (bcd) were used. The reduction steps began with removing and replacing unstable sensitive pixels (rogue pixels) using the SSC tool irsclean. 
The next step was averaging the individual exposures for each module. 
Fringes were present in the LL1 module (20--24\,$\mu$m) in four objects, and were removed using the defringe tool in Smart. 
For the high-resolution modules, the spectra were extracted using full aperture extraction. 
For the low-resolution modules, the background was removed by subtracting the averaged intensities at two off-target nod-positions for a given module. 
The low-resolution spectra were extracted using tapered column extraction with a slit width of 4 pixels.
For ten faint objects, an advanced optimal extraction algorithm \citep{Lebouteiller:2010dw} was applied, though the results of these two extractions were 
nearly identical in the end. 
It is worth mentioning that five objects in the low-resolution observations exhibited either additional sources and/or complex structure within the slit. 
For four sources, manual optimal extraction was performed to properly isolate the sources and the background, while one source did not require this process.
For these five sources the background was removed by subtracting the slit-order position as opposed to the nod position.

\begin{figure*} 
 \rotatebox{90}{ \begin{minipage} {13cm} 
 \resizebox{\hsize}{!}{\includegraphics*[51,155][525 ,761]{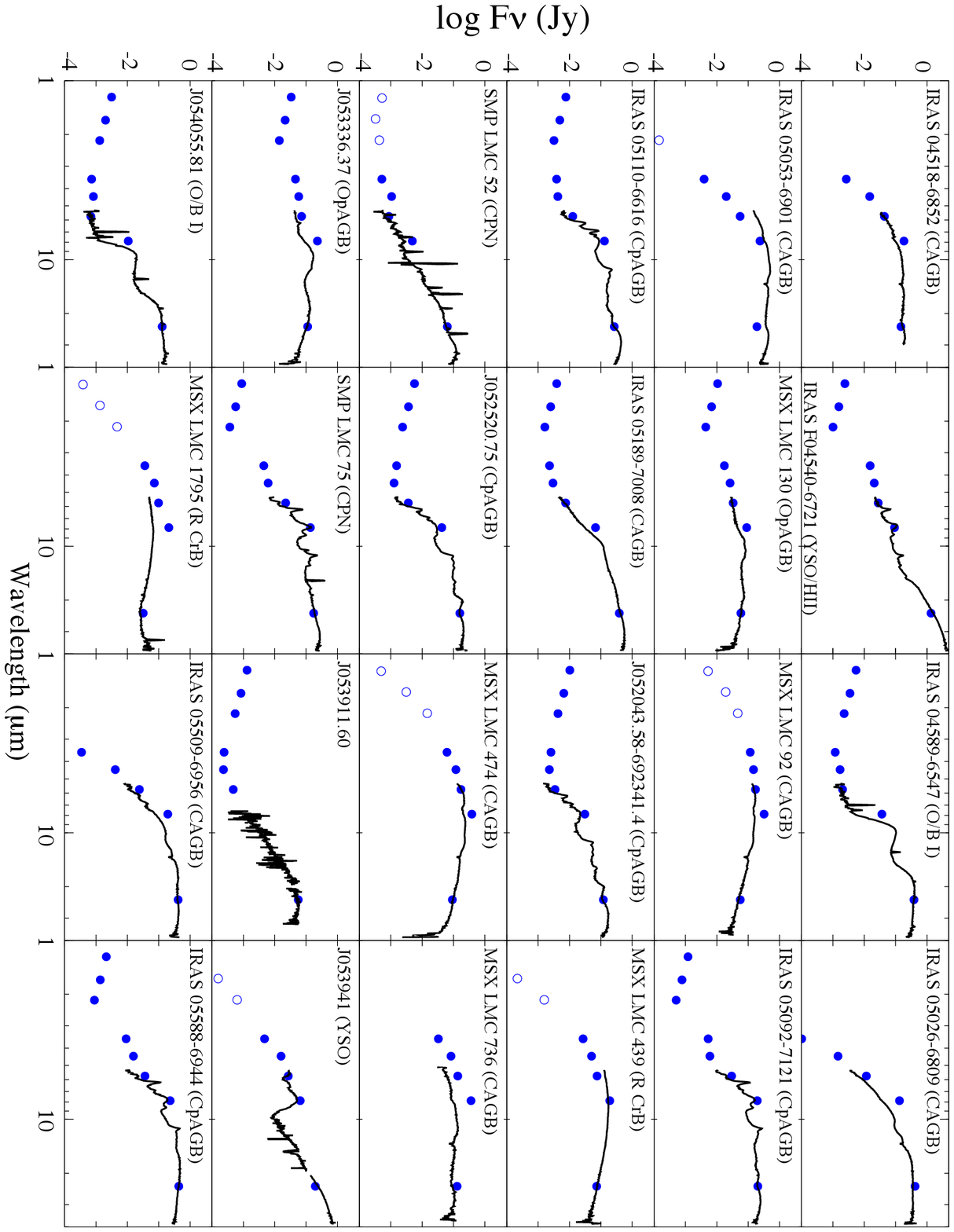}}  
 \end{minipage}}
  \caption{Infrared SEDs of our targets. {\it Spitzer} IRS spectra are plotted as black lines, accompanied by {\it Spitzer}-SAGE and 2MASS photometric points as 
blue filled circles. Photometric data from the IRSF  are plotted as open circles.
      \label{plot_sed}}
\end{figure*}

Figure\,\ref{plot_sed} shows the SEDs of our targets.
The SEDs were constructed using photometric data from 2MASS \citep{Skrutskie:2006hl}, the IRSF \citep{Kato:2007vx}, SAGE \citep{Meixner:2006eg} and 
our {\it Spitzer} IRS spectra.
Our primary targets were dust-embedded objects, whose SEDs peak at 10--20\,$\mu$m.
About half (13 objects) of the targets have double-peaked SEDs;
the first peak is at near-infrared wavelengths or shorter and the second peak at mid-infrared wavelengths.
The first peak likely corresponds to the stellar component, and the second peak corresponds to the thermal dust emission from the circumstellar envelope.
The low-resolution spectra of IRAS 04518$-$6852 and MSX\,LMC,736 were taken from the SAGE-Spec project \citep{Kemper:2010bwa}.

The 2MASS photometry for IRAS 05189$-$7008 and MSX\,LMC\,736 could not be used in Fig.\,\ref{plot_sed} because of contributions from nearby field stars that 
are coincidentally located close to the carbon-rich AGB stars  (Appendix.\,\ref{appendix-optical}).

 \citet{Gruendl:2009p27672} has listed IRAS 05189$-$7008 as a YSO, because this object
shows unusual colours and magnitudes as AGB stars do, and because its colour and magnitude could be closer represented by YSOs.
Our Spitzer IRS spectrum shows that this star is one of the reddest carbon stars with very weak molecular and dust features,
and IRAS 05189$-$7008 
pushes the limits of the carbon-rich AGB star boundaries pertaining to
the colour-magnitudes and colour-colour beyond the known range.

\section{Spectra and object classifications}

The {\it Spitzer} spectra of 24 targets were classified according to the object types. 
We found AGB stars, post-AGB stars, PNe, YSOs/H{\small II} regions and O/B supergiants.
Hereafter, we treated R\,CrB stars separately from post-AGB stars, because their spectra are different.
We describe the spectra and the reason for object classifications in the following subsections, and list object classifications in Table\,\ref{table-targets}.

\begin{center}
\begin{landscape}
\begin{table}
\begin{minipage} {23cm} 
  \caption{ List of targets  \label{table-targets}}
  \begin{tabular}{llrrrclllllcc}
\hline
no. & Object name & AOR & \multicolumn{2}{c}{SAGE Coordinates} & Selection$^\dag$ & Date$\ddag$ & Mode/ Exp. Time & Obj. Class.$^{\S}$ & Spt./Obj. Class$^{\P}$\\
& & & \multicolumn{2}{c}{(J2000)}  & & & & Our work & Literature\\
\hline
1  & IRAS\,04518$-$6852          & 27985664  &  04h51m40.56s  & $-$68d47m34.9s & SAGE & 04-29 & SH (120\,s$\times$6)                      & C AGB                 & C AGB$^{1}$\\
2  & IRAS\,F04540$-$6721         & 25996800  &  04h54m03.62s  & $-$67d16m18.2s & (iii) & 04-28 & SL (14\,s$\times$3), LL(6\,s$\times$3)   & YSO/HII               & Em$^{2}$\\
3  & IRAS\,04589$-$6547          & 25997056  &  04h59m07.39s  & $-$65d43m13.3s & (iii) & 04-28 & SL (60\,s$\times$3), LL(6\,s$\times$3)   & O/B supergiant        & O9ep/B1-2ep$^{3,2}$ \\
4  & IRAS\,05026$-$6809          & 25991680  &  05h02m31.47s  & $-$68d05m35.8s & (i)   & 04-28 & SL2 (60\,s$\times$3), SL1 (14\,s$\times$3), LL (6\,s$\times$3) & C AGB & C AGB$^{4}$, ERO$^{5}$ & \\
5  & IRAS\,05053$-$6901          & 25991168  &  05h05m04.81s  & $-$68d57m48.3s & (iv)  & 04-28 & SL (14\,s$\times$3), LL(14\,s$\times$3)  & C AGB \\
6  & MSX\,LMC\,130               & 25991936  &  05h07m18.33s  & $-$69d07m42.9s & (v)   & 04-28 & SL (14\,s$\times$3), LL(30\,s$\times$3)  & O post-AGB            & A7II/A9I$^{3}$, A7e$^{2}$ \\
7  & MSX\,LMC\,92                & 25990656  &  05h08m25.59s  & $-$68d53m59.9s & (v)   & 04-28 & SL (14\,s$\times$3), LL(30\,s$\times$3)  & C AGB                 & C$^{2}$ \\
8  & IRAS\,05092$-$7121          & 25992448  &  05h08m35.92s  & $-$71d17m30.7s & (i)   & 04-29 & SL (14\,s$\times$3), LL(14\,s$\times$3)  & C post-AGB            & late B--early G + H$\alpha$ em.$^{2}$ \\
9  & IRAS\,05110$-$6616          & 25992704  &  05h11m10.65s  & $-$66d12m53.7s & (i)   & 04-27 & SL (14\,s$\times$3), LL(14\,s$\times$3)   & C post-AGB           & F3II(e)$^{3}$ \\
10 & IRAS\,05189$-$7008          & 25996288  &  05h18m25.67s  & $-$70d05m32.6s & (iii) & 04-29 & SL (60\,s$\times$3), LL(6\,s$\times$3)   & C AGB                 &  \\
11 & J052043.58$-$692341.4       & 27985920  &  05h20m43.58s  & $-$69d23m41.4s & (ii)  & 04-29 & SL (60\,s$\times$3), LL(14\,s$\times$3)  & C post-AGB            & F5Ib(e)$^{3}$ \\
12 & MSX\,LMC\,439               & 25990400  &  05h20m48.21s  & $-$70d12m12.5s & (v)   & 04-28 & SL (14\,s$\times$3), LL(30\,s$\times$3)  & R CrB                 & R\,CrB$^{6}$ \\
13 & SMP\,LMC\,52                & 25996544  &  05h21m23.83s  & $-$68d35m33.5s & (iii) & 04-29 & SL (60\,s$\times$15), LL(30\,s$\times$3) & C PN                  & C PN$^{7,8,9}$ \\
14 & J052520.75$-$705007.3       & 25996032  &  05h25m20.75s  & $-$70d50m07.3s & (iii) & 04-28 & SL (60\,s$\times$3), LL(14\,s$\times$3)  & C post-AGB            & A1Ia$^{3}$, F2-F5I$^{2}$ \\
15 & MSX\,LMC\,474               & 25992192  &  05h25m51.85s  & $-$68d46m34.3s & (v)   & 04-29 & SL (14\,s$\times$3), LL(14\,s$\times$3)  & C AGB  \\
16 & MSX\,LMC\,736               & 25993216  &  05h33m06.80s  & $-$70d30m34.9s & (v)   & 04-28 & SH (120\,s$\times$6)                     & C AGB                 &  \\
17 & J053336.37$-$692312.7       & 27985408  &  05h33m36.07s  & $-$69d23m12.7s & (ii)  & 04-28 & SL (14\,s$\times$3), LL(14\,s$\times$3)  & O post-AGB            & G8pe$^{3}$ \\
18 & SMP\,LMC\,75                & 25993984  &  05h33m46.99s  & $-$68d36m44.2s & (iv)  & 04-29 & SL1 (14\,s$\times$3), LL (14\,s$\times$3), SH(120\,s$\times$10)  & C PN & PN$^{7,8,9}$\\
19 & J053911.60$-$693125.6       & 25995776  &  05h39m11.60s  & $-$69d31m25.6s & (iii) & 04-29 & SL1 (60\,s$\times$18), LL(30\,s$\times$4)& Unclassified \\
20 & J053941$-$692916            & 27986176  &  05h39m41.08s  & $-$69d29m16.7s & YSO   & 04-28 & SL (14\,s$\times$3), LL1(14\,s$\times$3), SH(120\,s$\times$10) & YSO & YSO$^{10}$ \\
21 & J054055.81$-$691614.6       & 25995520  &  05h40m55.81s  & $-$69d16m14.6s & (iii) & 04-29 & SL (60\,s$\times$15), LL(14\,s$\times$3) & O/B supergiant        & O--B$^{3}$, O--Bpe$^{2}$ \\
22 & MSX\,LMC\,1795              & 25990912  &  05h42m21.90s  & $-$69d02m59.1s & (iv)  & 04-28 & SL (14\,s$\times$3), LL(30\,s$\times$4)  & R\,CrB           & R\,CrB cand.$^{6}$, C$^{2}$  \\
23 & IRAS\,05509$-$6956          & 25991424  &  05h50m26.30s  & $-$69d56m02.8s & (i)   & 03-06 & SL (6\,s$\times$3),  LL(6\,s$\times$3)   & C AGB \\ 
24 & IRAS\,05588$-$6944          & 25992960  &  05h58m25.97s  & $-$69d44m25.3s & (i)   & 04-28 & SL (6\,s$\times$3), LL1(6\,s$\times$3), LL2(14\,s$\times$3)& C post-AGB   & WC?$^{3}$ \\
\hline
\end{tabular} \\
$^\dag$ the selection methods as described in Sect.\ref{targets}. SAGE: observed by the SAGE-spec project \citep{Kemper:2010bwa}. YSO: follow-up of AKARI YSO studies \citep{Shimonishi:2008du}.
$^\ddag$ Observed dates in 2009 and {\it mm-dd} format,
$^{\S}$ The object classifications based on our work, and 
$^{\P}$ spectral types or object types (R CrB/PNe) from the literature,
$^{1}$ \citet{Woods:2010it},
$^{2}$ this work,
$^{3}$ \citet{2011A&A...530A..90V},
$^{4}$ \citet{vanLoon:2006fr}, 
$^{5}$ \citet{Gruendl:2008fy},
$^{6}$ \citet{Soszynski:2009uaa},
$^{7}$ \citet{Jacoby:1980dw}, 
$^{8}$ \citet{Sanduleak:1978vj},
$^{9}$ \citet{Leisy:1996wi},
$^{10}$ \cite{Shimonishi:2008du}.
\end{minipage}
\end{table}
\end{landscape}
\end{center}


\begin{figure*}
\centering
  \resizebox{\hsize}{!}{\includegraphics*{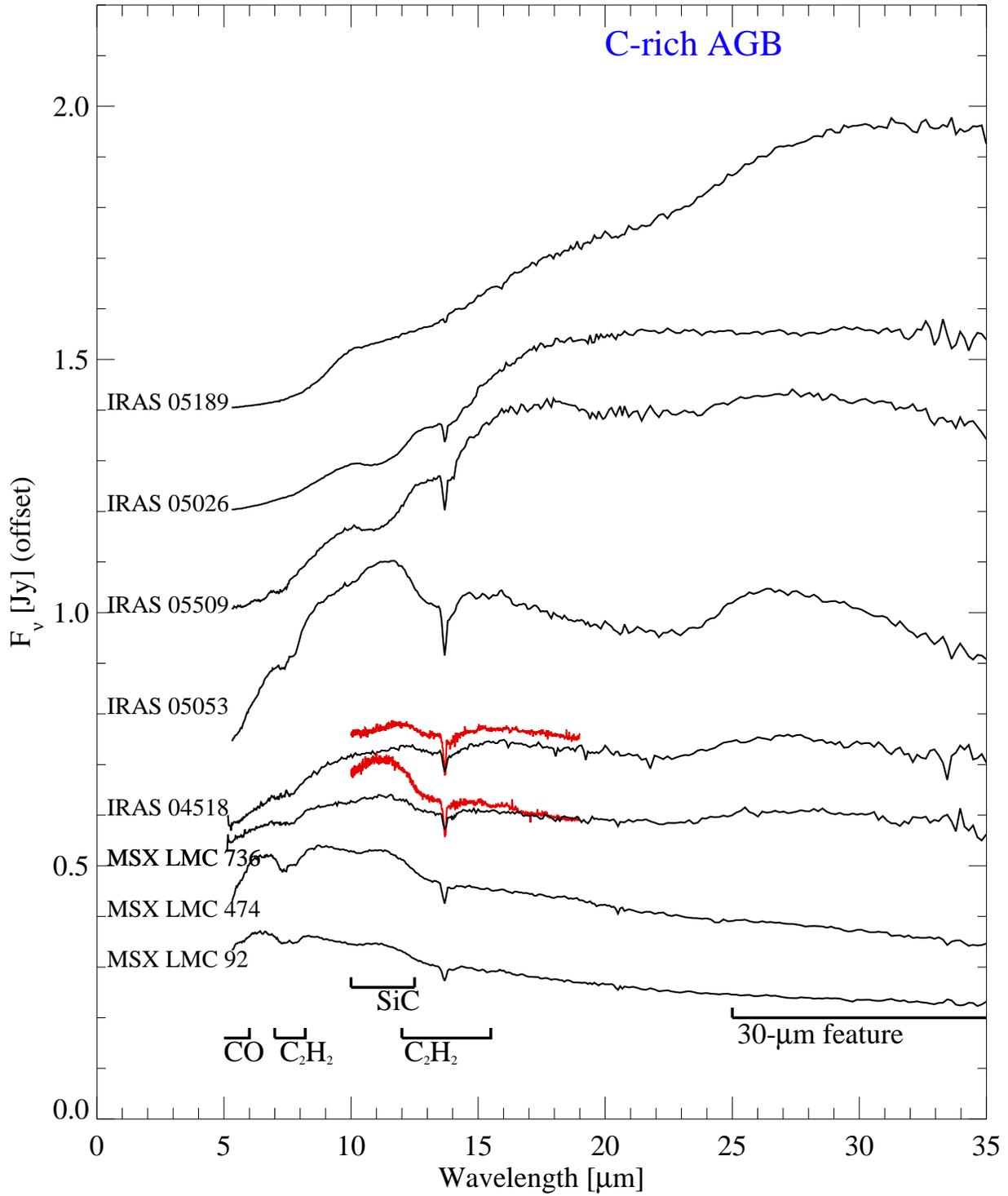}} 
   \caption{Spectra of C-rich AGB stars observed in the LMC. 
 The molecular (C$_2$H$_2$ and CO) and dust (SiC and the 30\,$\mu$m feature) features are indicated.
 Fluxes are offset by arbitrary amounts for clarity.
  High-resolution spectra of IRAS 04518$-$6852 and MSX\,LMC\,736 are plotted in red lines.
   The spectra are plotted in order of the underlying continuum temperature, from cooler (top) to warmer (bottom).
      \label{cagb}}
\end{figure*}

\subsection{Carbon-rich AGB stars}

Infrared spectra of carbon-rich AGB stars are characterised by C$_2$H$_2$ and HCN absorption bands at 7 and 14\,$\mu$m \citep[e.g.][]{Aoki:1999tm} together with SiC 
feature at 11.3\,$\mu$m \citep[e.g.][]{Zijlstra:2006gl}.
The SiC dust feature is normally found in emission, but has also been found in absorption in a few objects \citep{Gruendl:2008fy, Speck:2009fi}.
Some carbon-rich AGB stars show a very broad emission feature at 30\,$\mu$m \citep{Zijlstra:2006gl}, which is often attributed to MgS dust \citep{Hony:2001iw}, but the 
origin is still being debated \citep{Volk:2002gd, Messenger:2013bf}.
No PAHs or ionised lines are found, and the SEDs usually display a single peak.

Carbon-rich AGB stars are the most common objects among our targets, with 8 stars, as summarised in Table\,\ref{table-targets}.
This is a consequence of many factors. Stellar evolution models predict that the lifetime of post-AGB phase is shorter than that of AGB stars, 
so that more AGB stars are expected than post-AGB stars \citep{Vassiliadis:1993jk,Vassiliadis:1994p3049}.
The Large Magellanic Cloud has more carbon-rich AGB stars than oxygen-rich AGB stars
 \cite{Westerlund97}. 


Carbon-rich AGB spectra from our sample are shown in Fig.\,\ref{cagb}.
One can see the variation in the strengths of the C$_2$H$_2$ absorption features, which has a sharp Q-band branch feature at 13.7\,$\mu$m, accompanied with a broad wing 
formed by the P- and R-branches \citep{Matsuura:2006hx}.
The SiC feature, which appeared in emission in most of the objects, is particularly strong in IRAS 05053$-$6901, whereas the SiC feature was found in absorption in 
IRAS 05026$-$6809 and IRAS 05509$-$6956.
The spectrum of IRAS 05189$-$7008 has a weak SiC absorption.
There are six objects that show the broad 30\,$\mu$m feature (IRAS 05189$-$7008, IRAS 05026$-$6809, IRAS 05509$-$6956, IRAS 05053$-$6901, IRAS 04518$-$6852 and 
MSX\,LMC\,736).

In Fig.\,\ref{cagb}, the low-resolution spectrum of IRAS 04518$-$6852 was taken from the SAGE-Spec project \citep{Kemper:2010bwa}, and the high-resolution spectrum was 
from our programme.

\begin{figure*} 
\centering
  \resizebox{\hsize}{!}{\includegraphics*{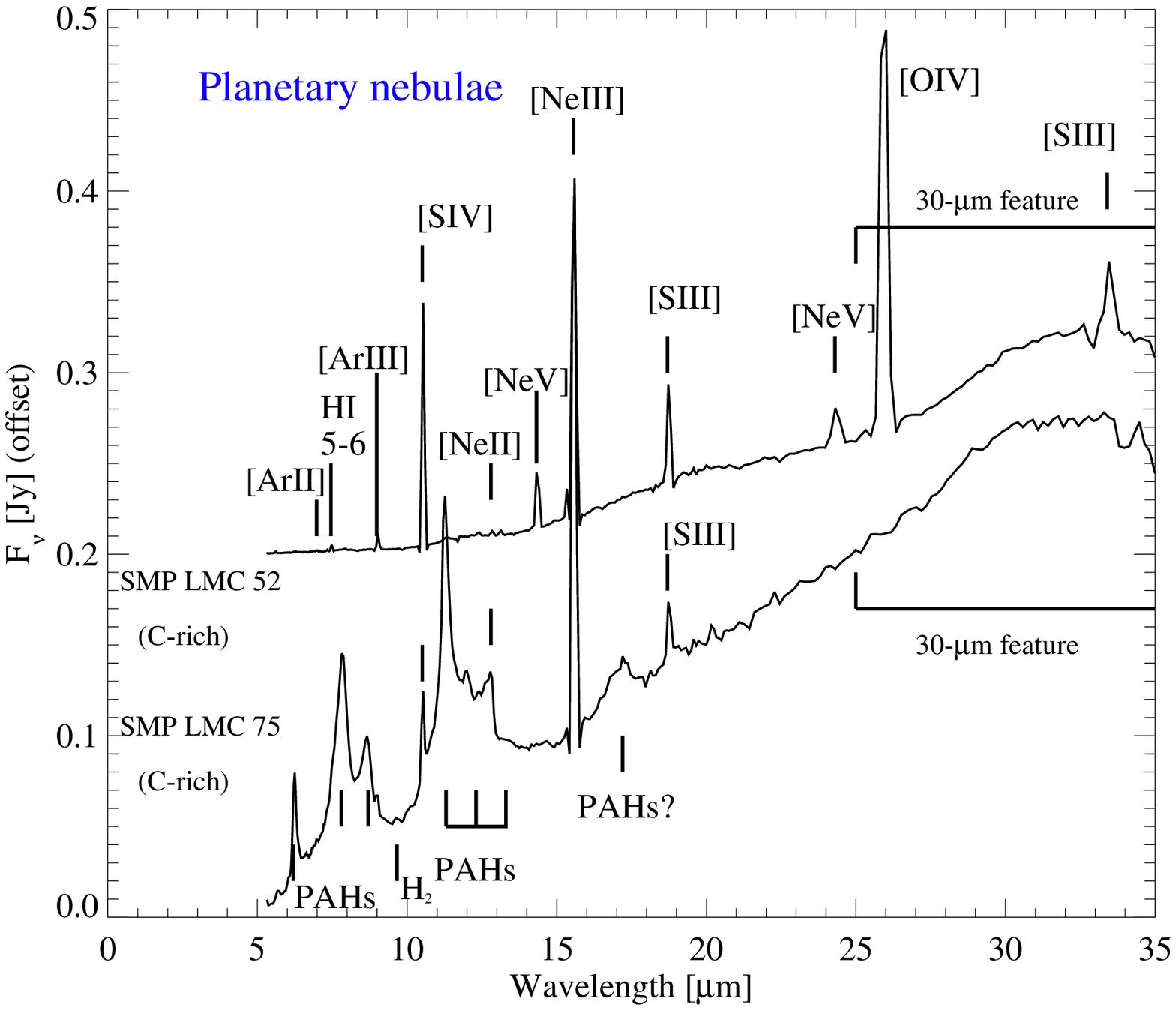}} 
   \caption{Spectra of the carbon-rich PNe observed in the LMC. 
   The spectrum of SMP\,LMC\,75 shows PAHs, indicating that this PN is carbon-rich.
   SMP\,LMC\,52 shows little sign of PAHs in the {\it Spitzer} spectrum, but the presence of the 30\,$\mu$m feature shows the carbon-rich nature of this star.
   \label{pn}}
\end{figure*}

\subsection{Carbon-rich PNe}

Carbon-rich PNe show emission lines, and lines detected include 6.99\,$\mu$m [Ar{\small II}], 12.81\,$\mu$m [Ne{\small II}]  and 33.48\,$\mu$m [S{\small III}] lines.
 A dust continuum is often found, but not always \citep{Stanghellini:2007kk, BernardSalas:2009iq}, and PAH features are also often found.
 SiC emission is occasionally found in LMC PNe, although it has rarely been found in Galactic PNe  \citep{Speck:2009fi, BernardSalas:2009iq}.

Our sample contains two C-rich PNe, SMP\,LMC\,52 and SMP\,LMC\,75; their spectra are shown in Fig.\,\ref{pn}.  
 The 30\,$\mu$m feature, which is associated with carbon-rich objects, is found in both spectra.
SMP\,LMC\,75 shows PAH features, including the one at 17.5\,$\mu$m \citep{Boersma:2010ci}.
 SMP\,LMC\,52 has only the 30\,$\mu$m feature as an indicator of carbon-rich nature, and it does not show PAHs. SMP\,LMC\,52 is a high-excitation PN, indicated by the 
detection of  [O{\small IV}] and [Ne{\small V}] lines \citep{BernardSalas:2009iq}.


\begin{figure*} 
  \resizebox{\hsize}{!}{\includegraphics*{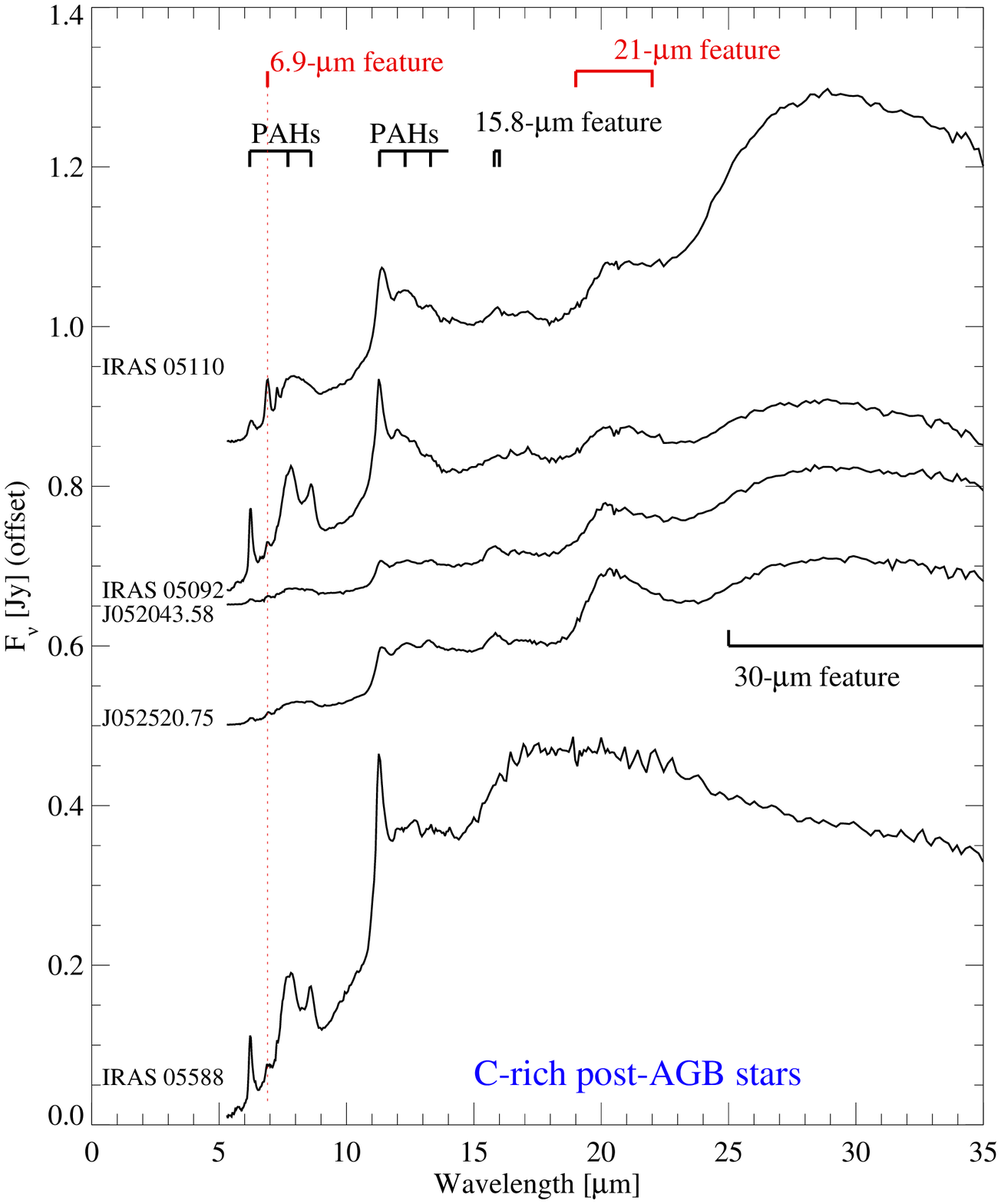}} 
   \caption{Spectra of carbon-rich post-AGB stars in the LMC.
   The approximate wavelengths of PAH features and four unidentified features (6.9, 15.8, 21 and 30\,$\mu$m features) are indicated.
     \label{cpagb}}
\end{figure*}

\begin{figure} 
  \resizebox{\hsize}{!}{\includegraphics*{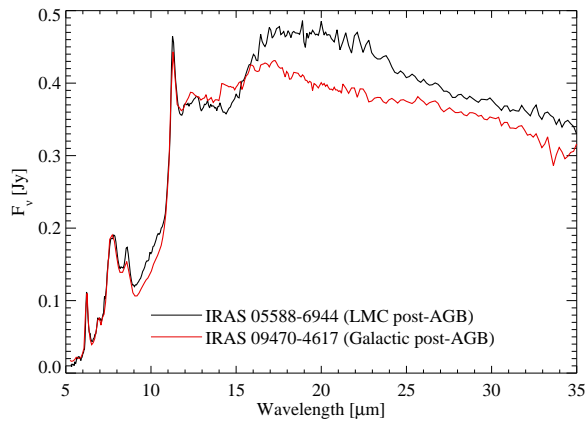}} 
   \caption{ Comparison of infrared spectra of a peculiar LMC post-AGB IRAS 05588$-$6944, with IRAS\,09470-4617.
   \label{Comp_IRAS05588_09470}}
\end{figure}

\subsection{Carbon-rich post-AGB stars} \label{sect-cpagb} 

As transitional objects from the AGB phase to the PN phase, carbon-rich post-AGB stars often show PAHs.


We classified five targets (IRAS 05092$-$7121, IRAS 05110$-$6616, J052043.58$-$692341.4, J052520.75$-$705007.3 and IRAS 05588$-$6944)
as carbon-rich post-AGB stars, and their spectra are presented in Fig.\,\ref{cpagb}.  
These spectra show PAHs, and  four out of the five spectra exhibit both the 21 and 30\,$\mu$m features.  
IRAS 05588$-$6944 is an exception, showing a rapid rise of the continuum up to 16\,$\mu$m and then a gradual decline, with perhaps the 30\,$\mu$m feature but no 
21\,$\mu$m feature.
An unidentified feature at 15.8\,$\mu$m, which may be associated with PAHs \citep{Moutou:1995ww, VanKerckhoven:2000vg}, also appears in some post-AGB stars.

The unidentified ``21\,$\mu$m feature'' has been found only in carbon-rich post-AGB stars.
 So far its detection has been reported in  16 Galactic stars \citep{Hrivnak:2009dg, Cerrigone:2011hv}, and a similar number (in total 11) of ``21\,$\mu$m feature'' objects have been found 
in the Magellanic Clouds \citep{Volk:2011iw}. \citet{Hrivnak:2009dg} suggested that the 21\,$\mu$m feature is related with carbonaceous molecules or dust,
while \citet{Li:2013th} proposed that it could be due to FeO. The identification of this feature is not yet well settled.

IRAS 05588$-$6944 is a peculiar object in our sample in several respects.
\citet{2011A&A...530A..90V} studied the optical spectrum and the SED of IRAS 05588$-$6944 as part of the LMC post-AGB study, and assigned IRAS 05588$-$6944 to be a 
WC (Carbon sequence Wolf-Rayet) candidate.
They estimated the luminosity of IRAS 05588$-$6944 to be 13\,000\,$L_{\odot}$, which is within the post-AGB luminosity limit 
\citep[35,000 $L_{\sun}$; ][]{2011A&A...530A..90V}, but the optical spectral classification of WC did not fit the post-AGB criteria.
Their figure C.2 showed the optical spectrum of IRAS 05588$-$6944, showing He{\small I} and C{\small II} lines but no hydrogen recombination lines.
The presence of carbon lines but a lack of hydrogen lines might suggest a WC nature.
However, the optical spectrum is not typical of WC stars with a circumstellar envelope, which are usually classified as WC9 or WC10.
\cite{2011A&A...530A..90V} pointed out that its optical spectrum lacked 5696\,$\AA$ C{\small III}, which is commonly found in WC10 stars \citep{Crowther:1998ec}. 
The infrared spectra of Galactic WC9 stars do not show PAHs \citep{vanderHucht:1996tn, Smith:2001bl}
but that of IRAS 05588$-$6944 shows strong PAH features (Fig.\ref{cpagb}).
No recombination lines nor ionised lines have been detected in the infrared 
spectrum of IRAS 05588$-$6944, even though such lines, indicative of an ionised nebula, have been found in infrared WC spectra
\citep{vanderHucht:1996tn, Smith:2001bl}.
The Spitzer spectrum IRAS 05588$-$6944 is compared with that of the Galactic post-AGB star IRAS\,09470-4617, which is scaled by a factor of 0.11
in Fig.\ref{Comp_IRAS05588_09470}. The Spitzer spectrum of IRAS\,09470-4617 is from \citet{Cerrigone:2009kn}, and they also have optical spectra.
 These two spectra look similar including the overall continuum, PAH features between 5--9\,$\mu$m
 and the lack of 21\,$\mu$m. Both show C{\small II} in the optical spectra.
It seems that  IRAS 05588$-$6944 and   IRAS\,09470-4617 could belong to a distinct class with a carbon-rich shell.

\begin{figure*} 
  \resizebox{\hsize}{!}{\includegraphics*{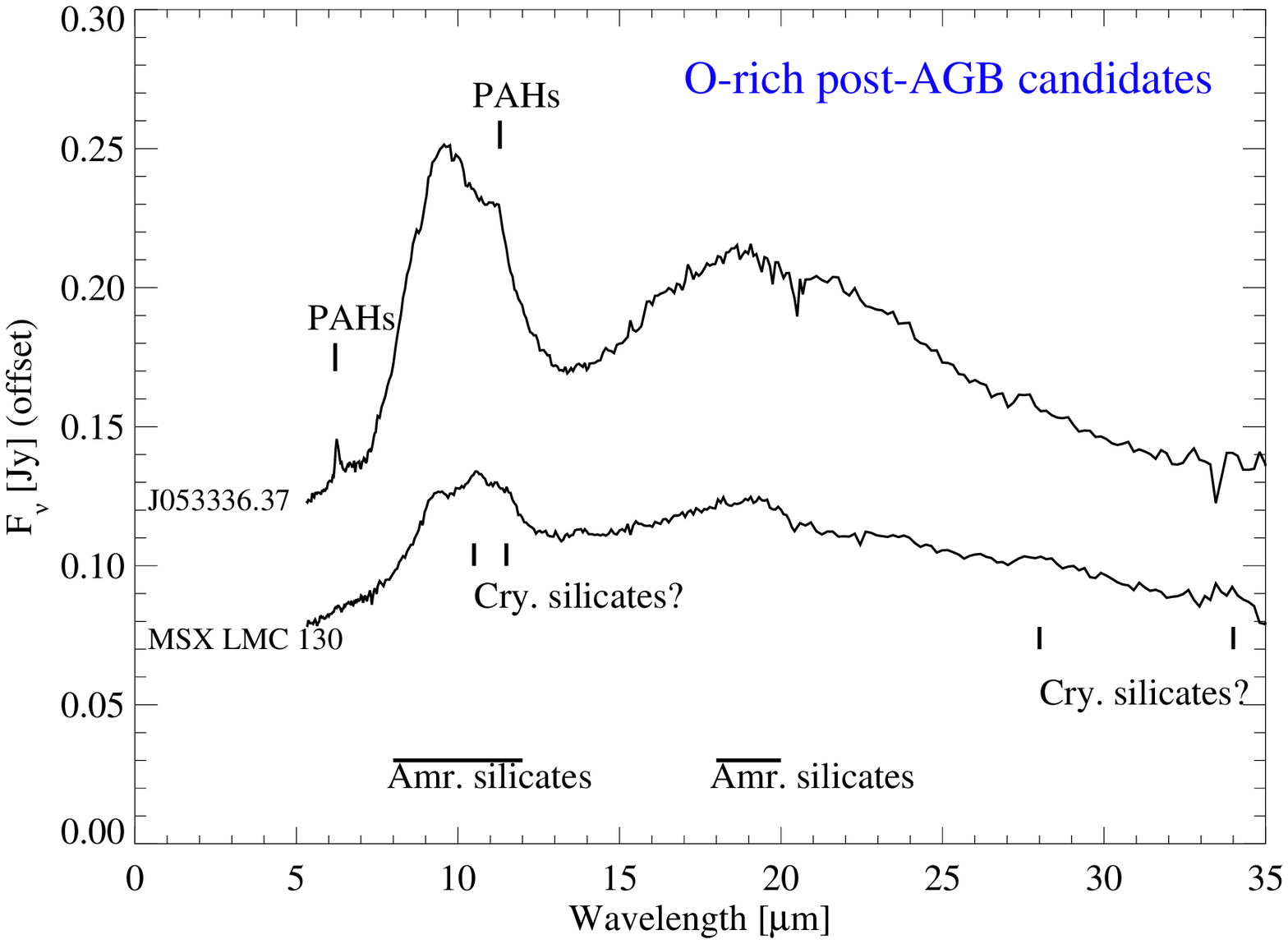}} 
   \caption{Oxygen-rich post-AGB candidates. Fluxes were offset by arbitrary amounts for clarity.
  Amorphous (Amr.) and potentially crystalline (Cry.) silicates were detected. PAHs in the spectrum of J053336.37$-$692312.7 are associated with the star itself 
rather than originating in the ISM.
   \label{opagb}}
\end{figure*}

\subsection{Oxygen-rich post-AGB stars} 

We found two oxygen-rich objects among our targets, and they are most likely post-AGB stars, because the spectral types of the central stars are A and G 
(Table\,\ref{table-targets}), and because their SEDs are double-peaked (Fig.\,\ref{plot_sed}).

Fig.\,\ref{opagb} shows spectra of two oxygen-rich stars.
Both of them show amorphous silicates, and MSX\,LMC\,130 shows hints of crystalline silicates.


The PAH features detected in the spectrum of J053336.37$-$692312.7 are likely to be associated with the star itself.
Along the IRS slit, PAH emissions are found only from the location of this star.
J053336.37$-$692312.7 is probably a mixed-chemistry object, which has both oxygen-rich and carbon-rich molecules and dust in a single star.
\citet{1998Natur.391..868W} proposed that a circumbinary disc may be responsible for some cases of mixed chemistry:
oxygen-rich dust had been ejected by the AGB star in the past and stored in the binary disc, while more recent mass loss was carbon-rich.
\citet{2011A&A...530A..90V} suggested that the SED of J053336.37$-$692312.7 can be modelled well as a post-AGB star with a circumbinary disc. It seems possible 
that mixed chemistry in this object is related to the presence of the disc.
Galactic post-AGB stars with mixed chemistry tend to have silicates with weak PAHs, and often crystalline silicates were more prominent than amorphous silicates
\citep{Molster:2002dh, Matsuura:2004fp, Gielen:2011ev, PereaCalderon:2009fk}.
In contrast, LMC post-AGB stars with mixed chemistry tend to show stronger amorphous silicate features than crystalline silicate ones, or even no indication of 
crystalline silicates \citep{Buchanan:2009gi}.
\citet{Jones:2013ih} proposed that metallicity differences might affect the crystalline silicate compositions, and one could assume that the detection rate of 
crystalline silicates could also be affected by metallicity.

\begin{figure} 
\centering
  \resizebox{\hsize}{!}{\includegraphics*{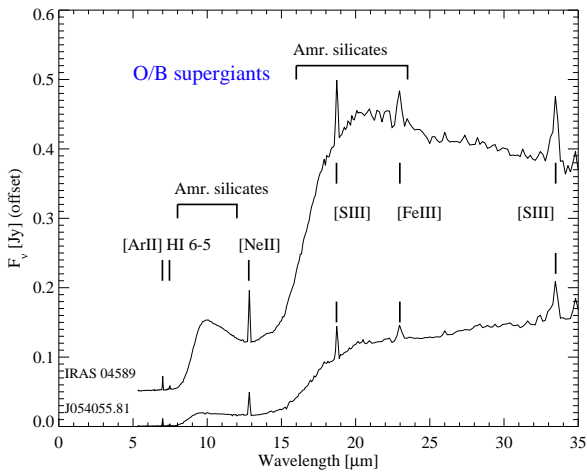}} 
   \caption{Spectra of the O/B supergiants with circumstellar envelopes. 
   Strong amorphous silicate emission and ionised lines were detected.
   \label{Be}}
\end{figure}

\subsection{O/B supergiants}

J054055.81$-$691614.6  and IRAS 04589$-$6547 are late O or early B-type stars \citep{2011A&A...530A..90V}. We discuss their optical spectra in more detail in 
Appendix\,\ref{appendix-optical}. 
The Spitzer spectra of these two stars show a series of ionised lines, including forbidden lines, and strong silicate emission bands at 10 and 18\,$\mu$m 
(Fig.\,\ref{Be}), indicating the presence of dust in their circumstellar envelopes.
The mid-infrared spectra of J054055.81$-$691614.6  and IRAS 04589$-$6547 resemble those of B[e] stars \citep{Voors:1999df, Kastner:2006hsa}
and luminous blue variables \citep[LBVs; ][]{Morris:1999du, Voors:2000wu};
both types of star have spectra with strong silicate emission bands and ionised lines.
However, these two O/B supergiants do not appear to be either B[e] stars or LBVs.
B[e]-type stars have forbidden emission lines of [FeII] and [OI] in their optical spectra and circumstellar dust emission in their mid-infrared spectra 
\citep{Allen:1976wj, Lamers:1998vh}. 
However, the optical spectra of J054055.81$-$691614.6 and IRAS 04589$-$6547 do not show either [FeII] or [OI] lines, so that they cannot be classified as B[e]-type stars.
 Furthermore, the luminosities of IRAS 04589$-$6547 and J054055.81$-$691614.6 were estimated to be 60\,000 and 70\,000\,$L_{\odot}$, respectively, 
\citep{2011A&A...530A..90V}, corresponding to about $M_{\rm bol}\sim-7$.
These luminosities are not sufficiently high to be LBVs, which have typical luminosities of $-8.5$--$-11$ $M_{\rm bol}$  \citep{Humphreys:1994fc}.
Thus, IRAS 04589$-$6547 and J054055.81$-$691614.6 appear to be
O/B supergiants that have hot circumstellar envelopes, but do not match known O/B supergiant or hypergiant categories with circumstellar envelopes, 
such as B[e]-type stars and LBVs.


\begin{figure} 
  \resizebox{\hsize}{!}{\includegraphics*{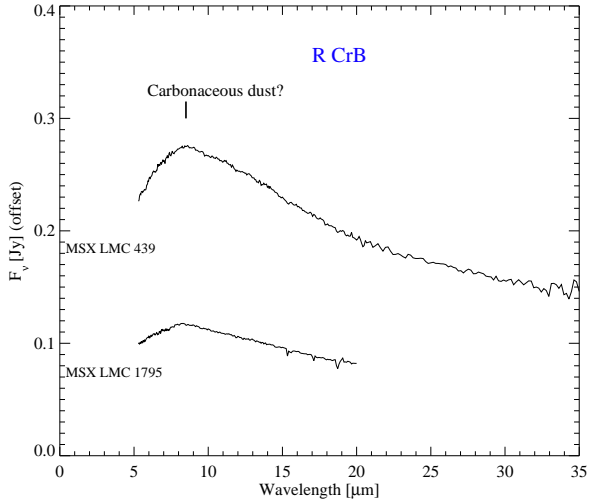}} 
   \caption{   
   Spectra of the R\,CrB stars, MSX\,LMC\,439 and MSX\,LMC\,1795. 
   Thermal emission from circumstellar dust is apparent.
   \label{rcrb}}
\end{figure}

\subsection{R CrB stars}

Two stars in our sample are an R\,CrB star and an R\,CrB candidate.
MSX\,LMC\,439 and MSX\,LMC\,1795 were classified as R\,CrB candidates, based on the analysis of MACHO optical light curves \citep{2001ASSL..265...71W}. 
OGLE light curves and follow-up optical spectroscopic studies confirm the R\,CrB status of MSX\,LMC\,439 but optical spectroscopic confirmation was not obtained 
for MSX\,LMC\,1795 \citep{Tisserand:2009kba, Soszynski:2009uaa}.
The optical spectrum of MSX\,LMC\,1795, discussed in  Appendix C, confirms that it is an R\,CrB star.
The {\it Spitzer} spectra of these two stars are shown in Fig.\,\ref{rcrb}.
The spectra, representing emission from circumstellar dust, are almost featureless with a very weak peak at 8--9 $\mu$m, like those of Galactic R\,CrB stars 
\citep{Lambert:2001cl, Clayton:2011ee}.

\begin{figure} 
  \resizebox{\hsize}{!}{\includegraphics*{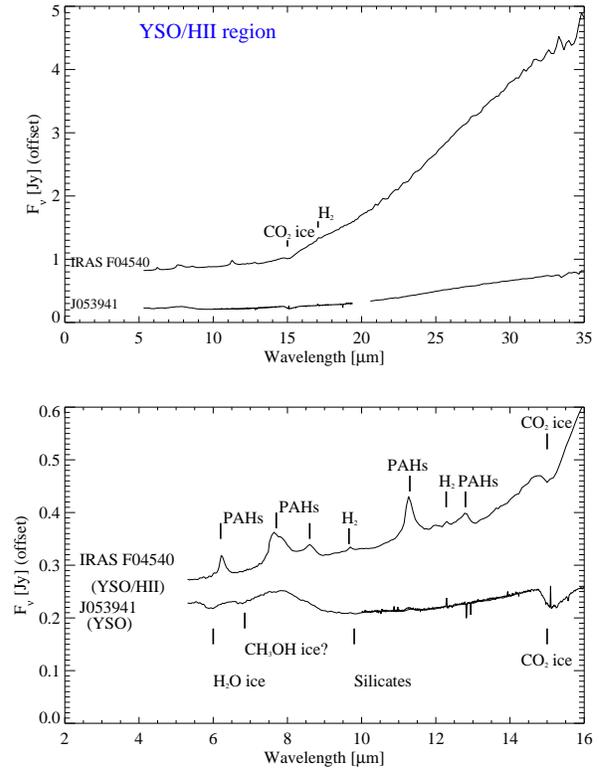}} 
   \caption{Spectra of the YSO and the HII region  observed in the LMC. 
   The spectral region between 6--16\,$\mu$m is enlarged in the lower panel.   
       \label{yso}}
\end{figure}

\subsection{YSOs/HII regions} 

Figure\,\ref{yso} shows the spectra of these objects, which are not evolved stars.
The spectrum of J053941$-$692916 shows CO$_2$ ice at 15.2\,$\mu$m, and the 4\,$\mu$m CO$_2$ ice band was also reported before \citep{Shimonishi:2010ek}, indicating 
that this object is a YSO \citep{Furlan:2006bb, Oliveira:2009es}.
Its rising SED up to 35 $\mu$m (Fig.\,\ref{plot_sed}) is also consistent with the known properties of a YSO.
The infrared spectrum and the SED resemble those of YSOs of Type-I \citep{Furlan:2008kp}.
The absorption at $\sim$6.8\,$\mu$m is probably associated with a CH$_3$OH ice band.

The {\it Spitzer} spectrum of IRAS F04540$-$6721 shows CO$_2$ ice at 15.2\,$\mu$m (Fig.\,\ref{yso}), but it also shows additional PAHs and H$_2$ emission.
The optical spectrum (Fig.\,\ref{spec_optical2}) shows hydrogen recombination lines, so there is ionisation.
The object appears to be a YSO, surrounded by a compact HII region. We classify this object as a YSO/HII region.





\section{ Colour-colour and colour-magnitude diagrams} \label{cc-diagrams}

IR colour-colour diagrams and colour-magnitude diagrams have been 
used to classify objects detected by photometric surveys \citep[e.g.][]{Blum:2006ib, Matsuura:2009fs}.
Such diagrams were also used to find post-AGB star candidates \citep[e.g.][]{2011A&A...530A..90V}.
As we used IR colours as one of the primary methods to select post-AGB candidates, we re-visit the post-AGB selection criteria on 
colour-colour and colour-magnitude diagrams.

We used five different types of colour-colour and colour-magnitude diagrams for the selection of the targets; we present one diagram in the main text, while 
the remaining diagrams are presented in Appendix\,\ref{appendix-cc}.
Overall these diagrams can find post-AGB candidates, but there is some contamination by other types of objects which have similar colours.

For comparison with the infrared colours of post-AGB stars and other types of objects, we incorporated existing spectroscopically known objects.
The sample assembled by \citet{Matsuura:2009fs}  is used, containing carbon-rich and oxygen-rich AGB stars and red-supergiants from the following sources:
\citet{Kontizas:2001ck},
\citet{Cioni:2001gq},
\citet{Sanduleak:1977wx},
\citet{Blanco:1980bu},
\citet{Westerlund:1981wi},
\citet{Wood:1983bh,Wood:1985jm},
\citet{Hughes:1989cr} and \citet{Reid:1988vj}.
We  further added samples of AGB stars, post-AGB stars, R\,CrB stars and PNe
from the SAGE-Spec project \citep{Kemper:2010bwa, Woods:2010it}.
Additionally, we assembled S\,Dor variables, a variety of the luminous blue variables (LBVs) from \citet{VanGenderen:2001kk},
WR stars from \citet{Breysacher:1999tx} and post-AGB stars from \citet{Gielen:2009iw}.
We removed probable galaxies prior to the analysis, following the colour selection used by \citet{Matsuura:2013js}.
We extracted photometric data from the {\it Spitzer}-SAGE LMC survey \citep{Meixner:2006eg}, which included near infrared photometry from 2MASS 
\citep{Skrutskie:2006hl}.

\subsection{Colour-colour diagram $K-[8.0]$ vs $K-[24]$}

\begin{figure*}
  \rotatebox{270}{ \begin{minipage} {13.2cm} 
  \resizebox{\hsize}{!}{\includegraphics*[75, 133][514, 706]{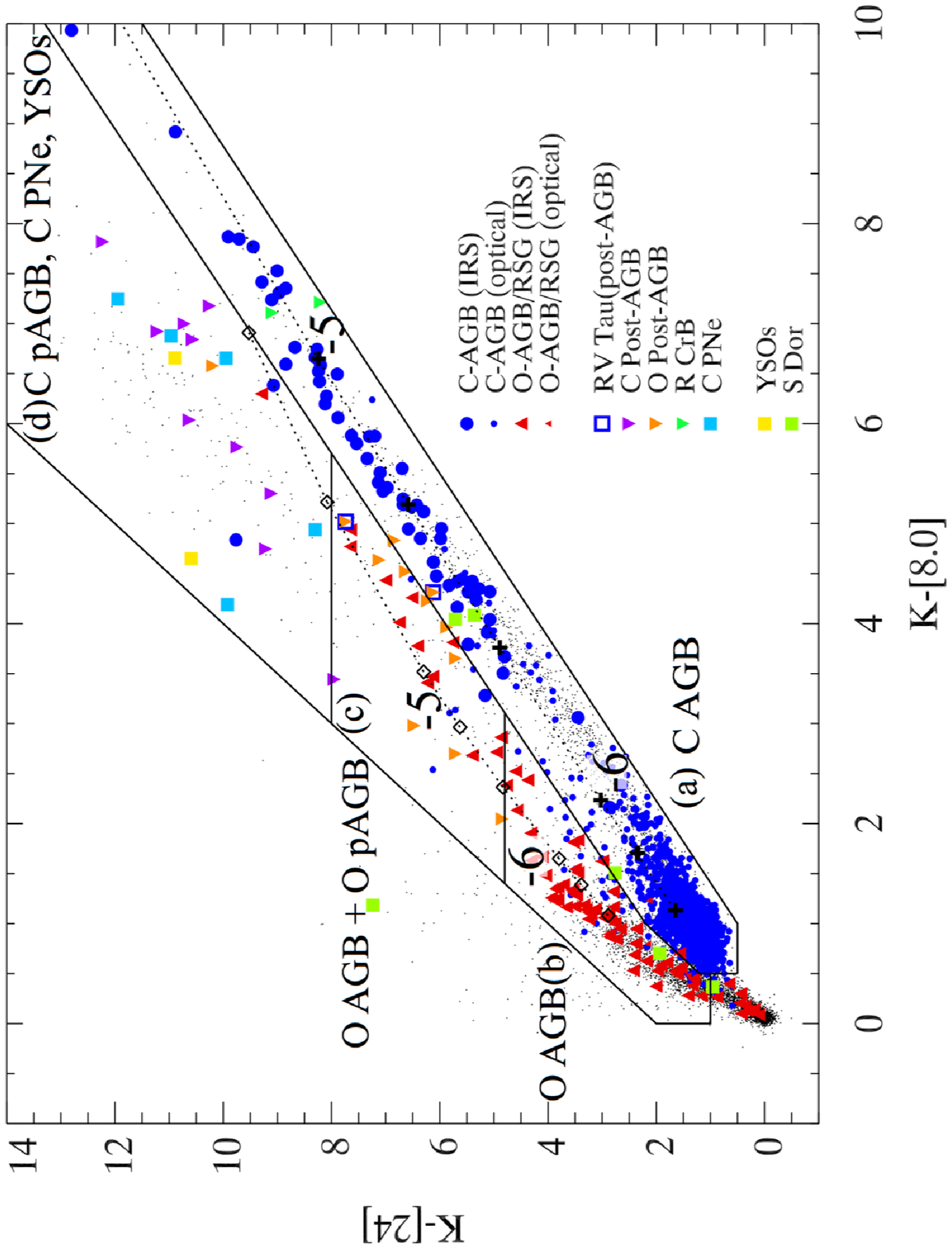}}  \end{minipage}} 
   \caption{The $Ks-[8.0]$ vs $Ks-[24]$ colour-colour diagram of the LMC objects.
     The black dots show all of the data in the SAGE database, combined with 2MASS photometric data. 
   Other symbols show the sources with known spectral classifications. C-AGB: carbon-rich AGB stars, where ``IRS" represents objects classified from {\it Spitzer} IRS 
observations, and ``optical" represents objects classified on the basis of optical spectra.
   O-AGB: oxygen-rich AGB stars, RV\,Tau: RV\,Tauri stars, C post-AGB: carbon-rich post-AGB stars, O post-AGB: oxygen-rich post-AGB stars, R\,CrB: R\,CrB stars, 
C PNe: carbon-rich PNe.
   The boxes highlight the object types typically found in the colour ranges indicated.
   Carbon-rich post-AGB stars are found in region (d) and are well-separated from carbon-rich AGB stars, but region (d) also contains PNe and YSOs.
   Oxygen-rich post-AGB stars are found in region (c).
    The two dotted lines show the typical colour at a given mass-loss rate of oxygen-rich (diamonds) and carbon-rich (plus signs) AGB stars.
   The diamond and plus sign symbols show the mass-loss rates $3\times10^{-7}$, $6\times10^{-7}$, $1\times10^{-6}$, $3\times10^{-6}$,
   $6\times10^{-6}$, $1\times10^{-5}$, $3\times10^{-5}$, $6\times10^{-5}$\,$\mlu$, and the colours for $1\times10^{-6}$ and $1\times10^{-5}$\,$\mlu$ are indicated 
   with $-6$ and $-5$.
   \label{cc_K8_K24}}

\end{figure*}

Figure\,\ref{cc_K8_K24} shows the colour-colour diagram of $K-[8.0]$ and $K-[24]$, which can be used to find post-AGB candidates.
The diagram contains slightly revised boundaries of regions from \citet{Matsuura:2013js}.
Oxygen-rich post-AGB stars are found mainly in region (c), with some in the redder part of region (b), whereas carbon-rich post-AGB stars are found in region (d).
In Fig.\,\ref{cc_K8_K24}, carbon-rich post-AGB stars are separated from carbon-rich AGB stars, because PAH features increase the [8.0]-band fluxes in 
carbon-rich post AGB stars.
Although the sample is small, there are an oxygen-rich AGB star and an oxygen-rich post-AGB star in region (d), as well as YSOs. 
Oxygen-rich post-AGB stars are mixed with high mass-loss rate oxygen-rich AGB stars in region (c).

\section{Analysis of PAH features}

The primary aim of our {\it Spitzer} observing programme was to investigate carbon-rich chemistry in LMC post-AGB stars.
More specifically, we aimed to understand if the shapes of PAH features would change along the path of stellar evolution from the AGB through the post-AGB to the 
planetary nebula phases, and if the low metallicity of the LMC has any impact on the PAH features.

In order to increase the sample size, we added spectra of post-AGB stars
that were observed by other {\it Spitzer} programmes, as was done in our previous paper \citep{Volk:2011iw}. 
The collected data are summarised in Table\,\ref{table-pagb}.
The data were assembled from programmes, proposal ID 30788 (P.I. Sahai)  and proposal ID 40159 
\citep[project name, SAGE-Spec; ][]{Kemper:2010bwa}.
As a part of SAGE-Spec, all LMC archival spectra were reduced and provided as a data product,
 we use these reduced data for this additional sample; the data reduction process was described by \citet{Kemper:2010bwa}.
Additionally, we included the HII region IRAS F04540$-$6721 and the PN SMP\,LMC\,75 from our {\it Spitzer} program in the PAH analysis, as both of them showed PAHs.

 \subsection{Continuum subtraction} \label{sect-PAH-cont}

\begin{figure} 
  \resizebox{\hsize}{!}{\includegraphics*{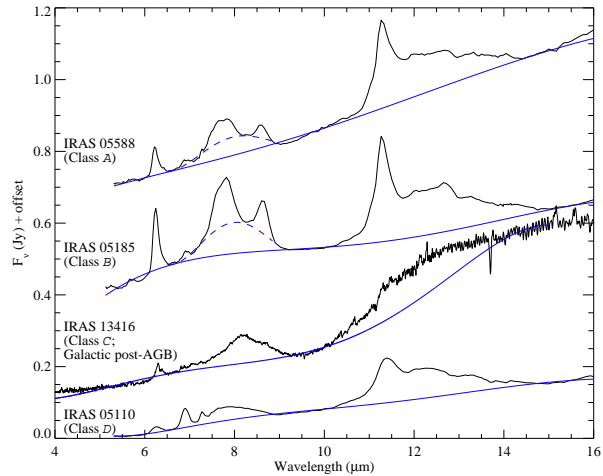}} 
   \caption{Examples of the continua (blue lines), which were subtracted from the observed spectra, in order to emphasise the intrinsic PAH spectra.
   The local continua at 6.5--9\,$\mu$m are indicated by dashed lines.
   \label{pah_cont}}
\end{figure}

In order to examine intrinsic profiles of PAH features, we subtracted the continua from the observed spectra.
 We estimated the continuum level by spline fits through the five  intervals (5.8--5.9,  6.8--7.0,  9.2--9.4,  9.8--10.2, and 14.3--15\,$\mu$m).
Some examples of continua are presented in Fig.\,\ref{pah_cont}. 
In addition, following \citeauthor{Peeters:2002ci}'s analysis of 6--9 $\mu$m features, we set local continua between 6 and 9\,$\mu$m on the spectra 
(Fig.\,\ref{pah_cont}) for class-$\mathcal{A}$ and class-$\mathcal{B}$, where classes will be discussed later.
This further subtraction of local continua makes it easier to examine the features at 7--9\,$\mu$m.
For 10--14\,$\mu$m, we did not set any local continuum, so that 
a `plateau', as described by \citet{Peeters:2004tl}, is included in our PAH spectral analysis if it is present.

\subsection{Features at 6--9\,$\mu$m}
 
\begin{figure*}
  \resizebox{0.8\hsize}{!}{\includegraphics*{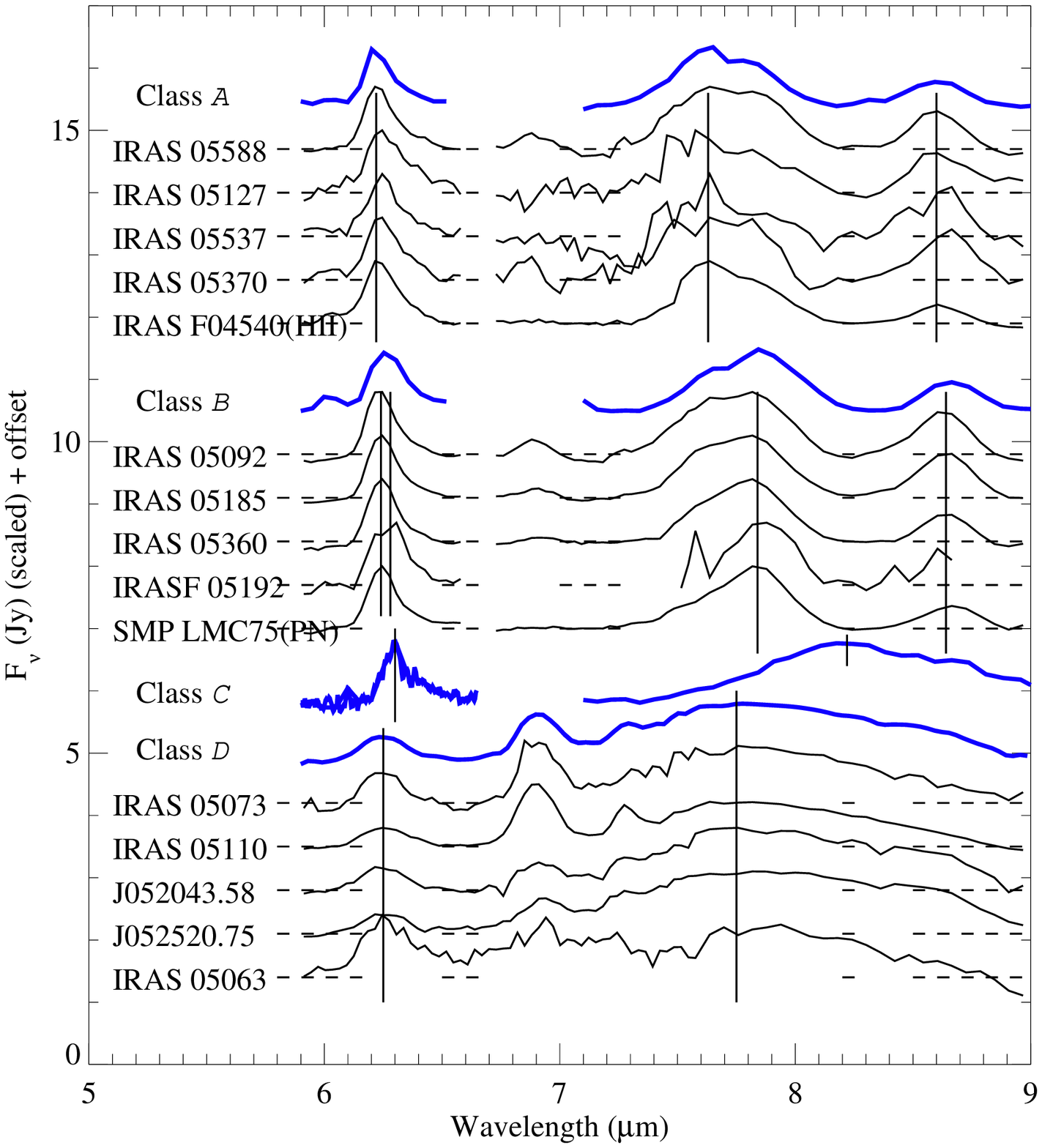}} 
   \caption{Continuum-subtracted {\it Spitzer} spectra at 6--9\,$\mu$m for C-rich post-AGB stars and two other objects in the LMC listed in Table 2. 
   Templates of  the class $\mathcal{A}$, $\mathcal{B}$ and $\mathcal{C}$ PAH profiles \citep{Peeters:2002ci} are plotted as bold lines, and PAH spectra of 
LMC post-AGB stars which match these classes follow below.
   Spectra of five LMC post-AGB stars do not match any previously known PAH profiles, and we name these as class $\mathcal{D}$. 
The average of these five class $\mathcal{D}$ spectra is plotted as the bold line.
   All of the spectra are scaled, and the `continuum' levels are marked with dashed lines.
   Vertical lines indicate the peak wavelengths of PAH features.
   \label{pah_6}}
\end{figure*}
\begin{figure*}
  \resizebox{\hsize}{!}{\includegraphics*{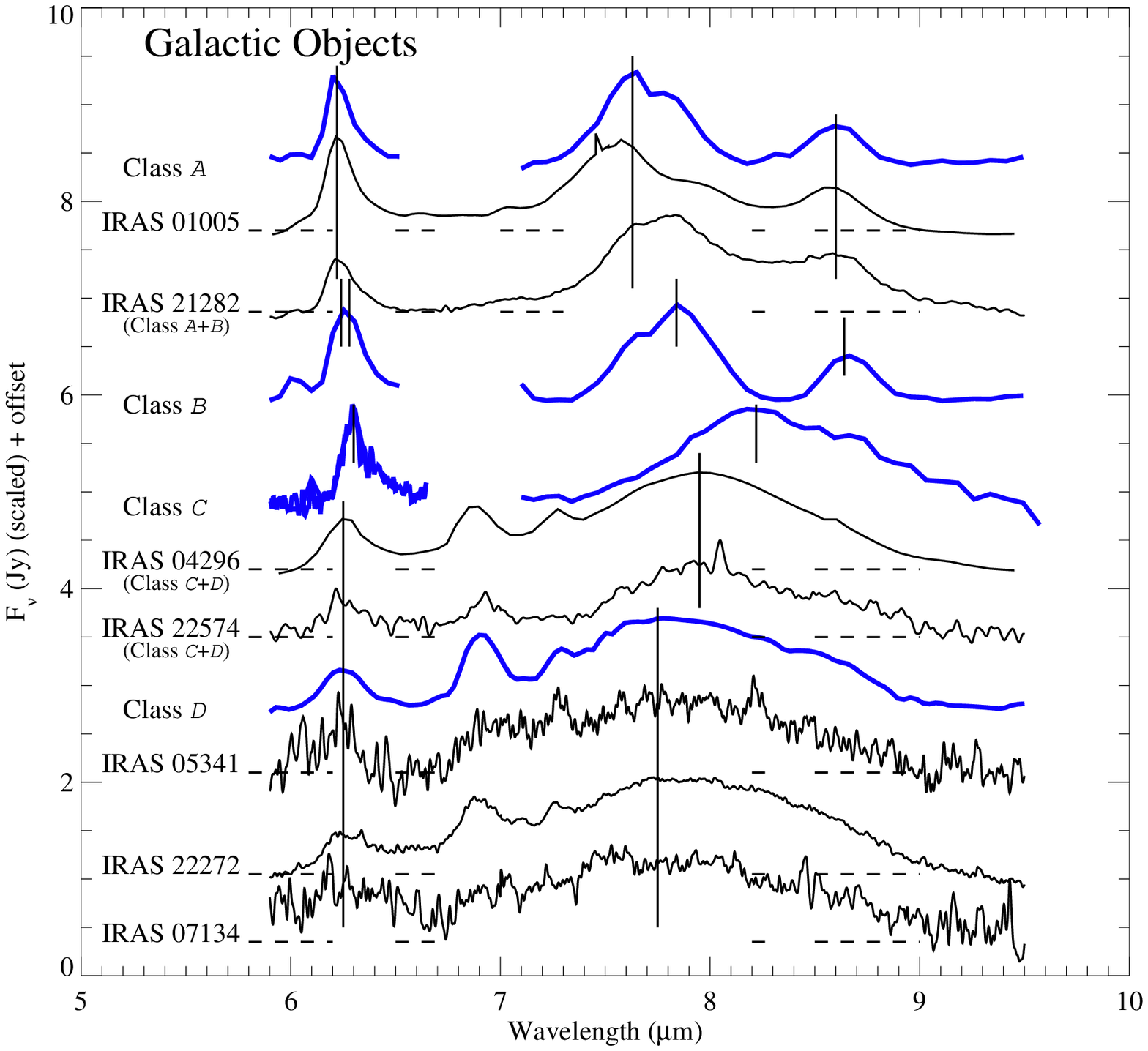}} 
   \caption{Continuum-subtracted spectra of representative Galactic post-AGB stars, showing the differences in the 6--9\,$\mu$m spectra.
   Template spectra of the classes $\mathcal{A}$--$\mathcal{C}$ and the average of the LMC class $\mathcal{D}$ profile are plotted.
   The spectra of the Galactic post-AGB stars, IRAS 04296$+$3429 and IRAS 22574$+$6609 show a profile between the classes $\mathcal{C}$ and $\mathcal{D}$, 
with a 7.5--9\,$\mu$m broad feature peaked at about 7.9\,$\mu$m.
   IRAS 05341$+$0852 and IRAS 22272$+$5435 show the class $\mathcal{D}$ profile.
   IRAS 22272$+$5435 has an additional unidentified feature at 6.35\,$\mu$m.
   \label{pah_6_Galactic}}
\end{figure*}

Figure \,\ref{pah_6} shows the 6--9\,$\mu$m continuum-subtracted spectra of 13 post-AGB stars, one PN and one YSO/HII region in the LMC.
Two or three features were found at $\sim$6.3, 7.6--8, and 8.6 $\mu$m, which are associated with PAHs.

\citet{Peeters:2002ci} examined the shapes of the features in the 6--8 $\mu$m region and divided the spectra into three classes according to the peak positions of 
the feature: $\mathcal{A}$, $\mathcal{B}$, and $\mathcal{C}$.
The templates of these profiles from \citet{Peeters:2002ci} are also plotted in Fig.\ref{pah_6}.
These templates were taken with the ISO/SWS, which had the spectral resolution of $R=\lambda / \Delta\lambda$=500--1500, and were smoothed with the 
{\it Spitzer} IRS spectral resolution.

In addition to broad features that are widely recognised as PAHs, there are features at $\sim$6.9\,$\mu$m and 7.3\,$\mu$m in some of the post-AGB stars (Fig.\ref{pah_6}).
Although the assignments of these two features are still debated \citep{Tielens:2008fx}, they are most likely attributed to aliphatic carbons 
\citep{Duley:1981we, Kwok:2001bc}.

\subsubsection{Class $\mathcal{A}$}

According to \citeauthor{Peeters:2002ci}'s scheme, class-$\mathcal{A}$ profile is characterised by peaks located at $\sim$6.22, 7.6 and $\sim$8.6\,$\mu$m, 
which correspond to subclasses A, A' and A'', respectively.
There is a secondary peak at about 7.8\,$\mu$m on the shoulder of the 7.6\,$\mu$m feature, but the 7.8\,$\mu$m peak is weaker than the 7.6\,$\mu$m peak.
The subclasses (A, A', A'') generally correspond to the major class $\mathcal{A}$ but some objects have slightly different subclasses from the major class 
\citep{Peeters:2002ci}.

The expected peak wavelengths of the profiles are marked with vertical lines in Fig.\,\ref{pah_6}.
 Amongst our sample, five objects have profiles matching the class $\mathcal{A}$ profile: four post-AGB stars, IRAS 05588$-$6944, IRAS 05127$-$6911, IRAS 05370$-$7019, 
IRAS 05537$-$7015 and the H{\small II} region IRAS F04540$-$6721.
Three objects (IRAS 05127$-$6911, IRAS 05370$-$7019 and IRAS 05537$-$7015) have noisy spectra, and we classified these three as class $\mathcal{A}$ based on 
their features having approximate peaks at about 7.6\,$\mu$m, but the peak at $\sim$6.2\,$\mu$m might suggest that these could be placed in class $\mathcal{B}$.

\citet{Peeters:2002ci} demonstrated that the class-$\mathcal{A}$ profile was found usually in objects heated by hot stars, whose spectral types are O-B. 
Objects with the class-$\mathcal{A}$ profile included H{\small II} regions and post-AGB stars.
Indeed, the H{\small II} region in our sample has a class $\mathcal{A}$ profile, which is consistent with \citeauthor{Peeters:2002ci}'s finding.

\subsubsection{Class $\mathcal{B}$}

The class $\mathcal{B}$ profile differs from that of class $\mathcal{A}$ by having the peaks of the three features shifted towards longer wavelength.
The peaks are found at about \,$\sim$6.24--6.28, 7.8--8.0 and $>$8.62\,$\mu$m, which have been assigned the subclasses B, B' and B'' \citep{Peeters:2002ci}.
Subclass B had a slightly wider feature around 6.2\,$\mu$m than subclass A.

Five objects have a class $\mathcal{B}$ (Fig.\,\ref{pah_6}) PAH profile:
four post-AGB stars (IRAS 05092$-$7121, IRAS 05185$-$6806, IRAS 05360$-$7121 and IRAS F05192$-$7008), and one PN (SMP\,LMC\,75).
All of these objects have the class B' feature peaking at about 7.8\,$\mu$m and the class B feature peaking at about 6.24\,$\mu$m.
The spectrum of IRAS F05192$-$7008 is noisy, but it seems to have three peaks corresponding to class $\mathcal{B}$.

\citet{Peeters:2002ci} found that objects with the class $\mathcal{B}$ profiles included PNe and post-AGB stars.
We found that the PN, SMP\,LMC\,75 is class $\mathcal{B}$, which
is consistent with what has been found for Galactic, and later for Magellanic Clouds PNe \citep{BernardSalas:2009iq}.

\subsubsection{Class $\mathcal{C}$}

  \citet{Peeters:2002ci} characterised class $\mathcal{C}$ spectra as having a peak at 6.3\,$\mu$m and a broad feature from 7.5--9.3\,$\mu$m, whose peak is 
located at 8.22\,$\mu$m  (Fig.\,\ref{pah_6}). 
Unlike classes $\mathcal{A}$ and $\mathcal{B}$, which have two separate features in the wavelength range of 7.5--9.3\,$\mu$m, class $\mathcal{C}$ has a 
single broad feature.

\citet{Peeters:2002ci} had a sample of 57 objects, but only two were assigned to class $\mathcal{C}$, both of which were post-AGB stars.
Later, more class $\mathcal{C}$ objects were found, including T Tauri stars and Herbig Ae and Be stars \citep{Keller:2008cj}.

In our sample, we did not find any objects matching the criteria of class $\mathcal{C}$, with a broad feature peaking at about 8.22\,$\mu$m.
Five objects in our LMC post-AGB sample show a similar profile to that of class $\mathcal{C}$ but 
the peak wavelength falls at a much shorter wavelength of about 7.7 $\mu$m.
In addition, their 6.25\,$\mu$m features are broader than those found in the Galactic class $\mathcal{C}$ objects,
and the peak is at 6.2 $\mu$m, shorter than the $\sim$6.3 $\mu$m for class $\mathcal{C}$. 
These five LMC objects are different from class $\mathcal{C}$, and we call their profile class $\mathcal{D}$.

The shift of the 8.2\,$\mu$m peak to 7.7\,$\mu$m is intrinsic to the object,
and is not an artefact of the continuum subtraction.
Figure\,\ref{pah_cont} shows the observed spectra, including the example of 
IRAS 05110$-$6616 (class $\mathcal{D}$) and IRAS 13416$-$6243 (Galactic
post-AGB star, class $\mathcal{C}$). 
The difference in the peak wavelength of the broad feature was already found in 
the observed spectra. 
This is further discussed in Appendix\,\ref{appendix-cont}.

\subsubsection{Class $\mathcal{D}$} \label{classD}

The five objects, mentioned above, which do not fit into the existing classes $\mathcal{A}$--$\mathcal{C}$, are taken as the prototypes of a new class $\mathcal{D}$.
Class $\mathcal{D}$ is characterised by a broader feature peaking at 6.24\,$\mu$m, and a broad single feature from 7--9\,$\mu$m, peaking at about 7.7\,$\mu$m.
The peak wavelengths are shifted shortwards, compared with class $\mathcal{C}$.
In Fig.\,\ref{pah_6}, five class $\mathcal{D}$ objects are plotted, and the average of these five class $\mathcal{D}$ profiles is plotted as a bold line.

 Class $\mathcal{D}$ has some additional differences from classes $\mathcal{A}$ and $\mathcal{B}$.
At 6.0\,$\mu$m, the class $\mathcal{A}$ and $\mathcal{B}$ templates have a small secondary peak next to the 6.2\,$\mu$m feature, 
but this feature is not apparent in the class $\mathcal{D}$ spectra.

All five class $\mathcal{D}$ objects in our LMC sample show an additional feature at 6.9\,$\mu$m.
This feature is not found or is weak in the class $\mathcal{A}$ and $\mathcal{B}$ objects in our sample: only a hint of this feature is found in IRAS 05092$-$7121 
and IRAS\,05185$-$6806.
Note that the templates from \citet{Peeters:2002ci} focused on the 6.3 and 7.2--9\,$\mu$m features, which are associated with PAHs,
so that this 6.9$\mu$m feature was not included in these templates (Fig.\,\ref{pah_6}).
The spectrum of the PN NGC 7027, which is type $\mathcal{B}$, shows the 6.9\,$\mu$m feature.
The origin of the 6.9\,$\mu$m feature is probably aliphatic carbon, but this has not been well established.

Among five class $\mathcal{D}$ objects in our sample, three stars (IRAS 05110$-$6616, J052043.58$-$692341.4 and 2MASS J052520.77$-$705007.5) have the 21\,$\mu$m feature 
and the other two  (IRAS 05063$-$6908 and IRAS 05073$-$6752) also have a hint of the 21\,$\mu$m feature \citep{Volk:2011iw}.
The class $\mathcal{D}$ and the 21\,$\mu$m feature may be of similar origin.

\citet{Peeters:2002ci} have pointed out that IRAS 22272+5435 and IRAS 07134+1005 show a broad single feature at $\sim$8\,$\mu$m similar to the class $\mathcal{C}$ 
profile, but its peak was shortwards of the class $\mathcal{C}$ 8.22\,$\mu$m peak. 
They did not include those two objects in class $\mathcal{C}$.
We classify these two stars as class $\mathcal{D}$ (Fig.\,\ref{pah_6_Galactic}). 
Additionally, IRAS 05341+0852 might potentially has the class $\mathcal{D}$ feature,
though its spectrum is too noisy to confirm this.
We searched for additional post-AGB spectra in the ISO/SWS archive (http://irsa.ipac.caltech.edu/data/SWS/)
and data collected by \citet{Sloan:2003ki}. 
We found that two Galactic post-AGB stars (IRAS 04296+3429 and IRAS 22574+6609) have a profile intermediate between those of classes $\mathcal{C}$ and $\mathcal{D}$.

 All the class $\mathcal{D}$ objects with known spectral types have a central star of type F or G (Tables\,\ref{table-pagb}, \ref{table-galactic}).
It is possible that spectral types of the central stars and PAH profiles are correlated.

\subsection{PAHs at 10--14$\mu$m}

\begin{figure*}
  \resizebox{0.7\hsize}{!}{\includegraphics*[45,31][497,752]{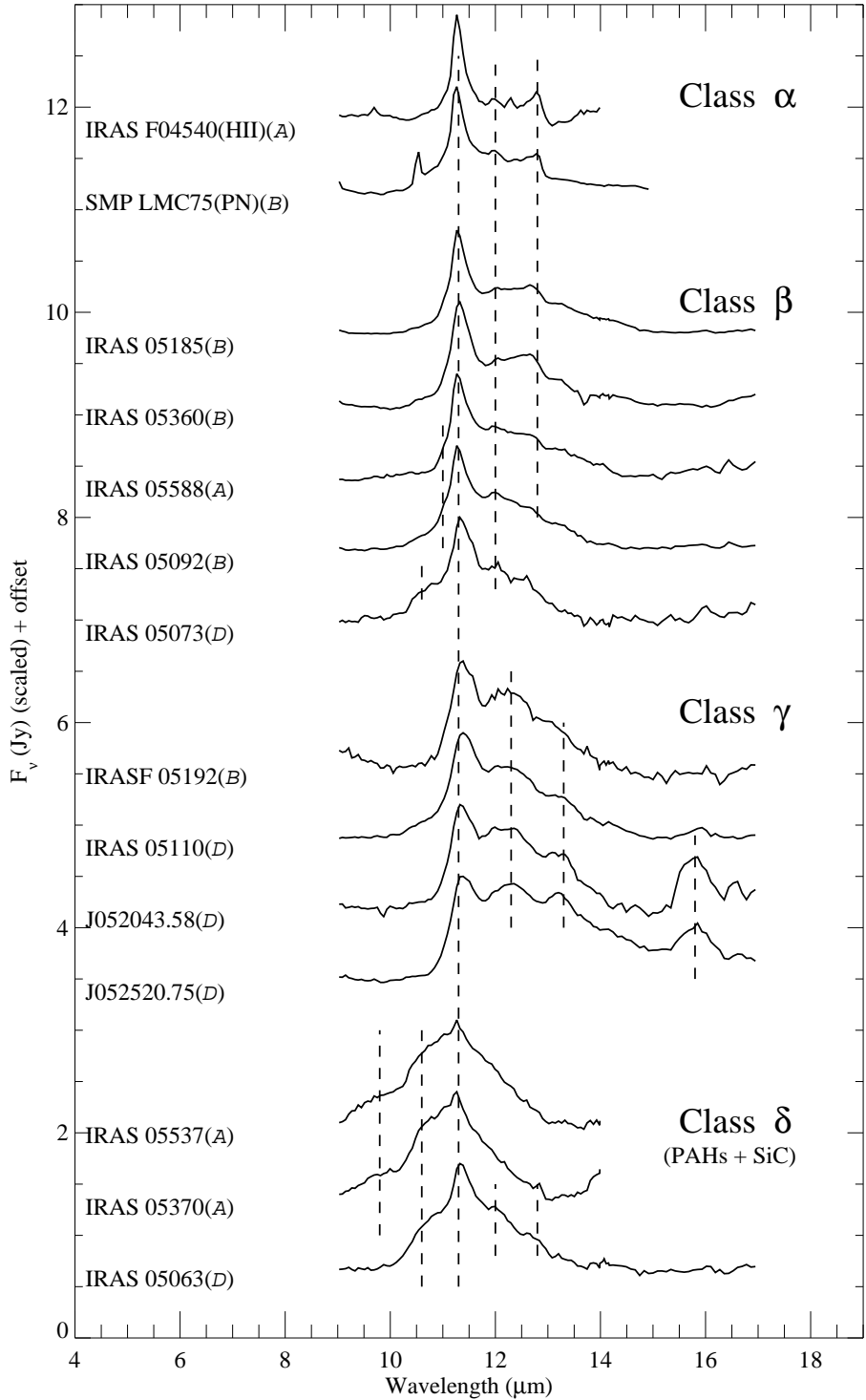}} 
   \caption{Continuum-subtracted {\it Spitzer} spectra at 9--17\,$\mu$m for C-rich post-AGB stars and two others in the LMC, from Table 2. 
The spectra are classified into four categories $\alpha$--$\delta$ based on the PAH features at 10--14\,$\mu$m
The class $\alpha$--$\gamma$ spectra are associated with PAHs, while the class $\delta$ profile is composed of PAHs and SiC.
   Object names are indicated on the left, followed by the 6--8\,$\mu$m PAH classes ($\mathcal{A}$--$\mathcal{D}$).
   \label{pah_11}}
\end{figure*}
\begin{figure}
  \resizebox{\hsize}{!}{\includegraphics*{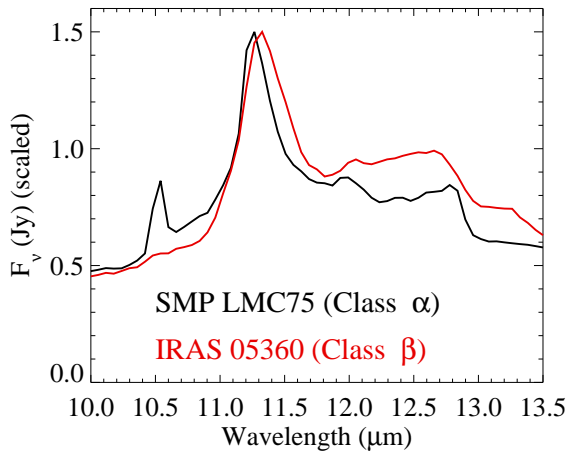}} 
   \caption{ Comparison between class $\alpha$ and $\beta$ features at PAH 10--14\,$\mu$m.
   Class $\beta$ has a wider 11.3\,$\mu$m feature, with a peak at longer wavelength, and the 12.7\,$\mu$m feature declines more gradually than in class $\alpha$.
   \label{pah_11_comp}}
\end{figure}
 
\begin{figure}
  \resizebox{\hsize}{!}{\includegraphics*{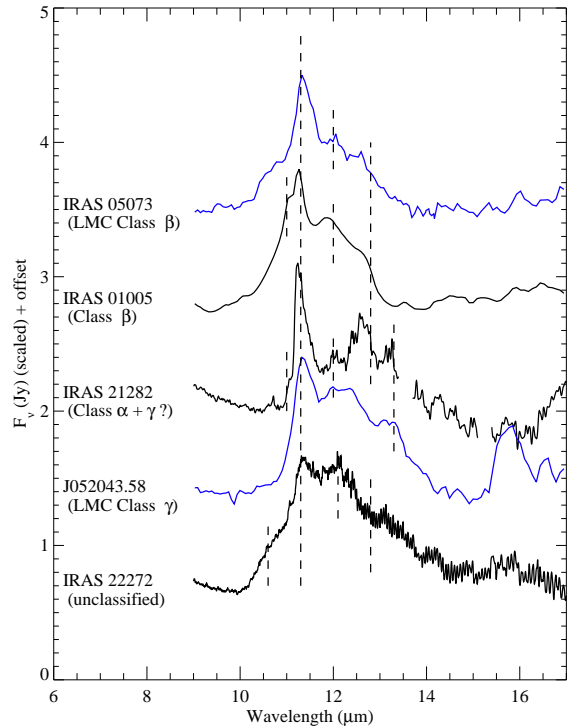}} 
   \caption{The 9--16\,$\mu$m continuum-subtracted spectra of Galactic post-AGB stars, whose 6--9\,$\mu$m spectra are shown in Fig.\,\ref{pah_6_Galactic}.
   Only spectra of good quality are plotted. The PAH classifications are indicated below the object names. The spectrum of IRAS 22272$+$5435 does not match 
any of the classes,  but possible SiC at 10.8\,$\mu$m suggests class $\delta$ with some contribution of PAHs at 13\,$\mu$m
as found in class $\gamma$ object.  \citet{Kwok:2001bc} proposed that the 9--16\,$\mu$m spectrum of IRAS 22272 might contain
aliphatic carbon features, making this spectrum unique.
   \label{pah_11_Galacticxs}}
\end{figure}

  The 10--14\,$\mu$m spectra of PAHs have been used as diagnostics for PAH substitution patterns \citep[e.g.][]{Socrates2001}.
Fig.\ref{pah_11} shows the continuum-subtracted 10--17\,$\mu$m spectra of fourteen objects in our sample.
The spectrum of IRAS\,05127$-$6911 was excluded, because it is too noisy in this wavelength range.

We grouped the PAH 10--14\,$\mu$m profiles  into four classes $\alpha$--$\delta$, based on the peak wavelengths and the shapes of the sub-features.
The characters of these classes are described in the following subsections, and the assigned classifications are summarised in Table\,\ref{table-pagb}.

For comparison, we included the classifications of the 6--8\,$\mu$m PAH features next to the object names in Fig.\,\ref{pah_11}.
There is no correlation between the classifications at 10--14\,$\mu$m and 6--9\,$\mu$m.

Previously, \citet{vanDiedenhoven:2004cw} classified the PAH profiles at 11.2\,$\mu$m into three categories A$_{11.2}$, A(B)$_{11.2}$ and B$_{11.2}$, as part of 
an overall PAH classification of 3--12\,$\mu$m spectra. 
However, because we found that the spectra at two different wavelength ranges of 10--14\,$\mu$m and 6--9\,$\mu$m are not correlated, we set up a separate classification 
scheme for 10--14\,$\mu$m.
We specifically use large math characters to classify the 6--9\,$\mu$m features and use $\mathcal{A}$ and $\mathcal{B}$ for the $\sim$6.2\,$\mu$m feature only, and 
we use Greek letters for 10--14\,$\mu$m.
\citeauthor{vanDiedenhoven:2004cw}'s  A$_{11.2}$ and B$_{11.2}$ categories relate to the 11.2\,$\mu$m feature only; the distinction is that the peak wavelength falls at 
11.2--11.24\,$\mu$m and 11.25\,$\mu$m, respectively, with FWHMs of 0.17 and 0.21, respectively.
We concluded that these characters fit our classes $\alpha$ and $\beta$, respectively, as defined below.
 \citeauthor{vanDiedenhoven:2004cw}'s third class, A(B)$_{11.2}$ has a feature at the same peak wavelength as A$_{11.2}$, but its width ($\sim$0.21\,$\mu$m in FWHM)
is slightly greater than that of class A$_{11.2}$, $\sim$0.17\,$\mu$m in FWHM.
B$_{11.2}$ and A(B)$_{11.2}$ can be distinguished with ISO/SWS spectra, which has a resolution ($\lambda/\Delta\lambda$) over 1000, but is hard to do so with the 
{\it Spitzer} IRS resolution (68--128). 
We do not set a classification equivalent to Class A(B)$_{11.2}$.

\subsubsection{Class $\alpha$}

The class $\alpha$ is characterised by a sharp 11.3\,$\mu$m feature, together with a well-isolated 12.7\,$\mu$m feature (Fig.\,\ref{pah_11}).
There is an additional feature at 12.0\,$\mu$m, but it is very weak.
The 11.3\,$\mu$m feature is the strongest among these three features, though in the Galactic sample, some objects have the 12.7\,$\mu$m feature as strong as 
the 11.3\,$\mu$m one \citep{Hony:2001iw}.
The combination of these three distinct features is typically found in PNe such as the Galactic PN NGC 7027 \citep{Hony:2001iw}.
In our sample, an H{\small II} region and a PN have this class.

\subsubsection{Class $\beta$}

 The class $\beta$ profile shows a slight difference from the class $\alpha$ (Fig.\,\ref{pah_11_comp}).
 The 11.3\,$\mu$m feature is wider, and the long wavelength edge of the feature declines much more gradually than in class $\alpha$.
The 12.7\,$\mu$m feature declines more gradually towards longer wavelength than in class $\alpha$.
The long wavelength tail of the 11--12\,$\mu$m complex can continue up to 14\,$\mu$m.
The feature at 12.0\,$\mu$m is also found but is often weak.
Among our LMC post-AGB sample, five objects have this type of profile, as does one Galactic post-AGB star (Fig.\,\ref{pah_11_Galacticxs}).

There is a very weak feature at 11.0\,$\mu$m on the shoulder of the 11.3\,$\mu$m bands in two objects (IRAS 05588$-$6944 and IRAS 05092$-$7121). 
This 11.0\,$\mu$m feature is also associated with PAHs \citep{vanDiedenhoven:2004cw, Tielens:2008fx}.
IRAS 05073 has a hint of SiC contribution, which is found in class $\delta$ objects.

\subsubsection{Class $\gamma$} \label{sec-15.8}

Spectra with broad triple features at 11.3, $\sim$12.3 and $\sim$13.3\,$\mu$m are classified as $\gamma$.
These features are broad and connected, rather than being two distinct features at 11.3\,$\mu$m and 12.7\,$\mu$m as found in the class $\alpha$ and $\beta$ objects.
The combination of features can sometimes extend to nearly 15\,$\mu$m, as found in 2MASS J05252077$-$7050075 (Fig.\,\ref{cpagb}).
Among our post-AGB sample, four objects have this profile.

Two of the class $\gamma$ objects have a feature at 15.8\,$\mu$m (J052043.58$-$692341.4 and 2MASS J05252077$-$7050075).
Its assignment has been suggested to be carbonaceous molecules or dust \citep{Hrivnak:2009dg, Volk:2011iw}.
There are PAH features at 15.8\,$\mu$m \citep{Socrates2001, Moutou:1996vs, VanKerckhoven:2000vg}, however, 
\citet{Hrivnak:2009dg} pointed out that the features found in post-AGB stars were wider ($\sim1.3$\,$\mu$m) than
the PAH features found in HII regions, PNe and Herbig AeBe stars.
C$_{70}$ has a feature at this wavelength as well \citep{Cami:2010ig},
but lack of other C$_{70}$ associated features at 17.5 and 18.7\,$\mu$m
eliminates this possibility.
The 15.8\,$\mu$m feature is probably associated with PAH ring deformation vibrations \citep{Socrates2001},
but additional contributions should be considered.

\subsubsection{Class $\delta$}

The remaining three objects are classified as class $\delta$.
 They have a very broad feature extending from 10 to 14\,$\mu$m (Fig.\,\ref{pah_11}), with a sharp PAH feature at 11.3\,$\mu$m on top.
 There is an additional weak feature at 10.8\,$\mu$m on the short-wavelength shoulder of the 11.3\,$\mu$m feature.
 The broad feature and the 10.8\,$\mu$m feature are associated with SiC, which we will discuss in the following section.
There is a broad secondary feature at 9.8\,$\mu$m, but its identification is unknown.

\subsection{Continuum and features}

\begin{figure}
  \resizebox{\hsize}{!}{\includegraphics*{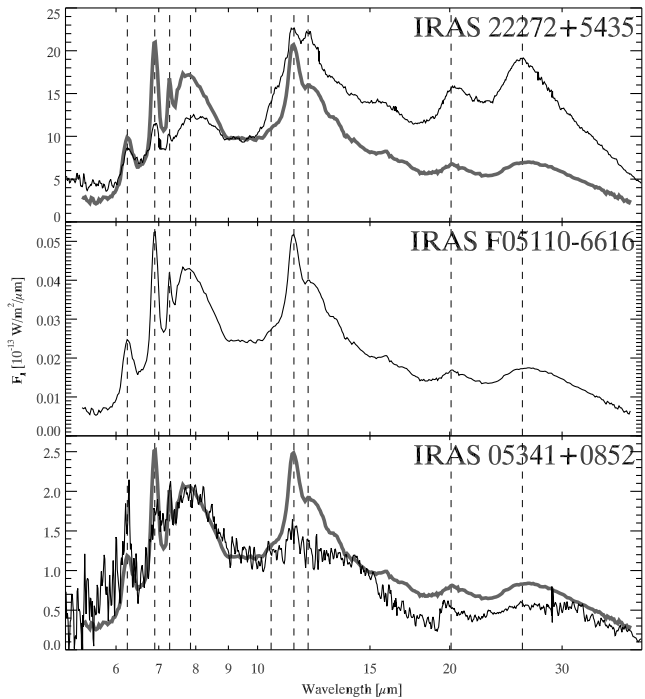}} 
   \caption{ Comparison of the spectra of the LMC post-AGB star, IRAS F05110$-$6616 with
    two Galactic post-AGB stars, IRAS 05341+0852 and IRAS 22272+5435.
    The top and middle panel, the spectrum of IRAS F05110$-$6616 is plotted in grey colour for comparison, with the flux scaled at 9.0--9.6\,$\mu$m.
    \label{Galactitc}}
\end{figure}

Figure.\ref{Galactitc} shows comparison of the overall PAHs and SEDs of the LMC post-AGB star,
IRAS F05110$-$6616 and two Galactic post-AGB stars, IRAS 05341+0852 and IRAS 22272+5435.
All three post-AGB stars have, or likely to have $\mathcal{D}$ PAH feature at 6--9\,$\mu$m,
IRAS F05110$-$6616  and IRAS 05341+0852 have similar spectra in this wavelength range, including the continuum levels.
In contrast, a difference of these two spectra is found at 10--14\,$\mu$m PAH features.
The difference at 10--14\,$\mu$m but similarities at 6--9\,$\mu$m PAHs and continuum suggest
that either compositions or bondings structures of PAHs that emit at two different wavelength
range differ.

\section{Fits with experimental PAH cross-sections}

\subsection{Methods}

We attempt to fit {\it Spitzer} spectra with laboratory-measured PAH cross-sections to test whether they, especially those of the new class $\gamma$, do indeed 
show the presence of PAHs. 
The features at 10--14\,$\mu$m were used to diagnose the structure of aromatic rings \citep{Socrates2001} and we focus on this wavelength region, 
which was extensively studied in the laboratory \citep[e.g.][]{Hudgins:1998ig}.

We fitted the {\it Spitzer} continuum-subtracted spectra at 10--14\,$\mu$m with a composite of the NASA Ames laboratory spectra \citep{Bauschlicher:2010gz, Mattioda10}.
The NASA Ames database contains theoretical and experimental cross sections, and we use the experimental cross sections, as they have higher accuracy in the wavelengths 
of the features.
We focused on PAHs consisting of carbon and hydrogen only, though the database includes PAHs containing nitrogen, oxygen and silicon.
That resulted in 40 laboratory spectra, consisting of 30 neutral and 10 ionised PAH compounds.
The smallest PAH compound used in the analysis was C$_{14}$H$_{10}$ (phenanthrene and anthracene) and the largest one was C$_{50}$H$_{22}$. 
The largest ionised compound was C$_{24}$H$_{12}^+$ (coronene).
The laboratory spectra were convolved with the Lorentzian line profile, and further smoothed to the {\it Spitzer} spectral resolution.
We tested the fits by minimising $\chi^2$.
We convolved the laboratory cross-section with a blackbody.
In the small spectral range of 10--14\,$\mu$m, the spectral gradient caused by the temperature variation was small.
 `Redshift' in the laboratory data \citep{Mattioda10} is ignored, as this shift is smaller than the Spitzer spectral resolution.
Some {\it Spitzer} spectra showed a trace of the SiC band at 11.3\,$\mu$m, so we included the SiC feature as well as PAHs in our $\chi^2$ fitting process. 
We extracted 10--15\,$\mu$m ISO/SWS spectra of the Galactic carbon star, W\,Ori \citep{Sloan:2003ki}, as a SiC template.

The spectrum of the PN, SMP\,LMC\,75, has an [S{\small IV}] emission line at 10.6\,$\mu$m and possibly the [Ne{\small II}]
line at 12.8\,$\mu$m. We masked the 10.6\,$\mu$m region,
but included the 12.8\,$\mu$m region, as the latter line was insignificant.

We attempted unsuccessfully to fit 6--9\,$\mu$m spectra in the same way.
This was partly hampered by the limitations of the $\chi^2$-minimisation fitting method.
The $\chi^2$-fitting tends to capture the global trend of the features, but sometimes has difficulty in simultaneously fitting a combination of broad and sharp features 
as found in observed 6--9\,$\mu$m spectra.
Additionally, post-AGB stars are likely to have both PAHs and aliphatic carbon features at this wavelength range, making it more difficult 
to fit \citep[e.g.][]{Kwok:2011iv, Hrivnak:2000dz}.
An alternative possibility is that nitrogen-substituted PAHs or hydrogenation might be important in this wavelength range \citep{Tielens:2008fx, Duley:1981we}.
We leave the fitting of the 6--9\,$\mu$m wavelength for future investigations.

\subsection{Results of PAH fitting}

\begin{figure}
  \resizebox{\hsize}{!}{\includegraphics*[22,24][498,764]{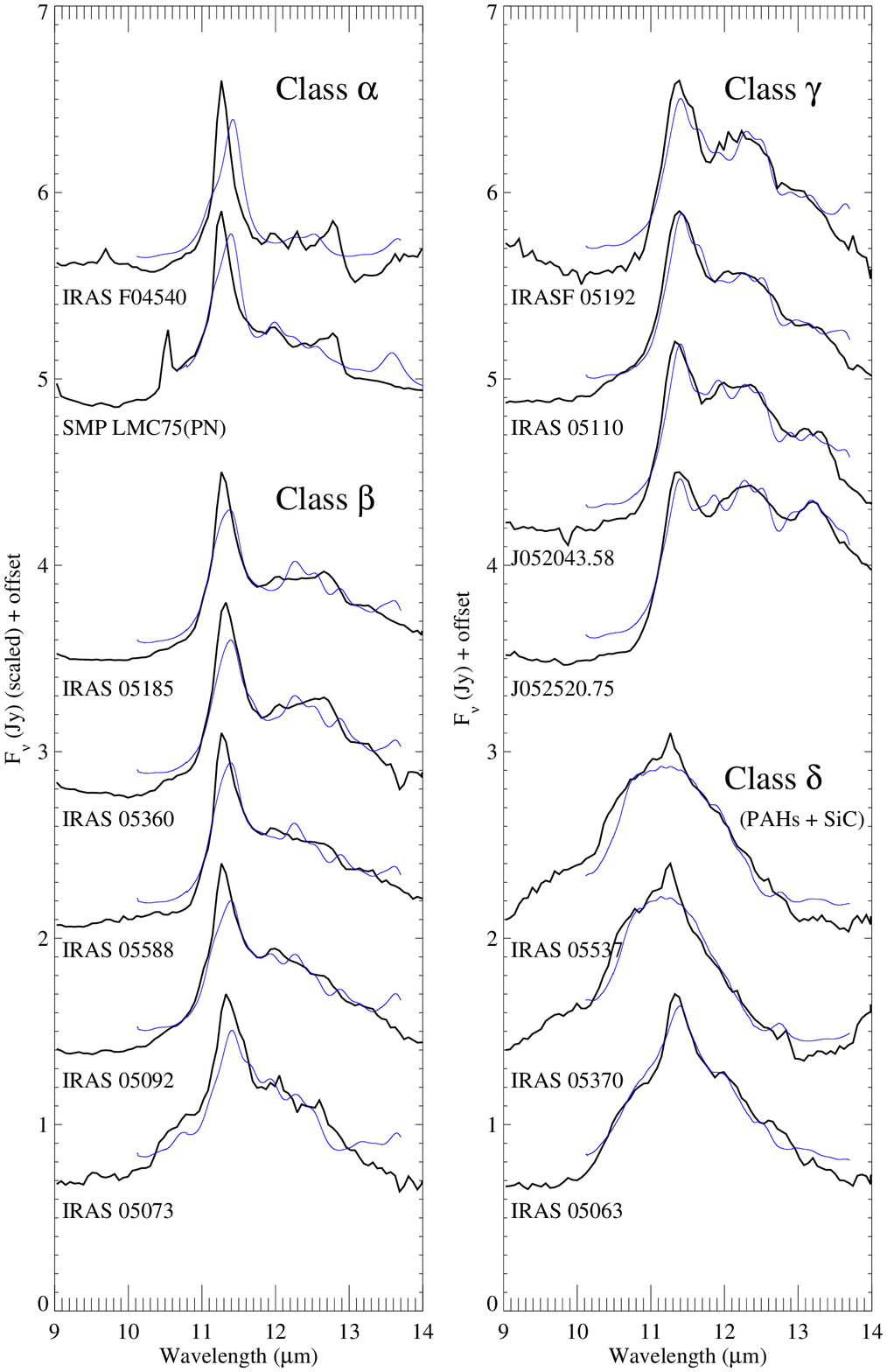}} 
   \caption{The composite PAH(+SiC) spectra (blue lines), compared with observed {\it Spitzer} spectra (black lines).
    \label{pah_lab_fit_spectra}}
\end{figure}

\begin{table*}
  \caption{List of carbon-rich post-AGB stars used for PAH analysis \label{table-pagb}}
\begin{tabular}{lllllllllc}
\hline
no & Names & SAGE Coordinates & Program ID  & Spt. & \multicolumn{5}{c}{PAHs}\\
 & & (J2000)        &                 & &    \multicolumn{4}{l}{6--9\,$\mu$m} & 11--14\,$\mu$m\\
 & & & &  & Overall$^{\dag}$ & 6.2 & 7.7$^{\ddag}$ & 8.6$^{\ddag}$ \\
\hline
2  & IRAS F04540$-$6721               & 04 54 03.62 $-$67 16 18.2 & 50338 & (YSO/HII)         & $\mathcal{A}$ & (A & A' & A'')                & $\alpha$   \\
   & IRAS 05063$-$6908                & 05 06 03.67 $-$69 03 58.8 & 30788 &                   & $\mathcal{D}$ & (D?& \multicolumn{2}{c}{D?)}  & $\delta$   \\
   & IRAS 05073$-$6752                & 05 07 13.91 $-$67 48 46.6 & 40159 &                   & $\mathcal{D}$ & (D & \multicolumn{2}{c}{D)}   & $\beta$    \\
8  & IRAS 05092$-$7121                & 05 08 35.94 $-$71 17 30.5 & 50338 & late B- early G   & $\mathcal{B}$ & (B & B' & A'')                &$\beta$     \\
9  & IRAS 05110$-$6616                & 05 11 10.61 $-$66 12 53.8 & 50338 & F3II(e)           & $\mathcal{D}$ & (D & \multicolumn{2}{c}{D)}   & $\gamma$   \\
   & IRAS 05127$-$6911                & 05 12 28.18 $-$69 07 55.7 & 40159 &                   & $\mathcal{A}$ & (B?& A'?& A'')                & (noisy)   \\
   & IRAS F05192$-$7008               & 05 18 45.26 $-$70 05 34.5 & 40159 &                   & $\mathcal{B}$? & (B?& B'?& ? )                 & $\gamma$  \\
   & IRAS 05185$-$6806                & 05 18 28.17 $-$68 04 04.1 & 30788 &                   & $\mathcal{B}$ & (B & B' & B'')                & $\beta$   \\
11 & J052043.58$-$692341.4            & 05 20 43.58 $-$69 23 41.4 & 50338 & F5Ib(e)           & $\mathcal{D}$ & (D & \multicolumn{2}{c}{D)}   & $\gamma$  \\
14 & 2MASS J05252077$-$7050075        & 05 25 20.75 $-$70 50 07.3 & 50338 & A1Ia/F2-F5I & $\mathcal{D}$ & (D & \multicolumn{2}{c}{D)}   & $\gamma$  \\
18 & SMP LMC 75                       & 05 33 46.96 $-$68 36 44.1 & 50338 & (PN)              & $\mathcal{B}$ & (B & B' & B'')                & $\alpha$  \\
   & IRAS 05360$-$7121                & 05 35 25.84 $-$71 19 56.7 & 30788 &                   & $\mathcal{B}$ & (B & B' & B'')                & $\beta$   \\
   & IRAS 05370$-$7019                & 05 36 32.52 $-$70 17 38.4 & 40159 &                   & $\mathcal{A?}$& (B?& A'?& B''?)               & $\delta$  \\
   & IRAS 05537$-$7015                & 05 53 11.96 $-$70 15 22.6 & 40159 &                   & $\mathcal{A}$ & (B?& A'?& A''?)               & $\delta$  \\
24 & IRAS 05588$-$6944                & 05 58 25.94 $-$69 44 25.7 & 50338 & WC?               & $\mathcal{A}$ & (A & A' & A'')                & $\beta$  \\
\hline
\end{tabular}\\
$^{\dag}$ The primary classification of the PAH features at 6--9\,$\mu$m, considering three features
at 6.2, 7.7 and 8.6\,$\mu$m. The Spitzer spectrum of
IRAS 05370-7019 is noisy and it is difficult to determine the PAH classifications.
$^{\ddag}$ For class D, there is one broad peak between 7--9\,$\mu$m rather than two distinct features
at 7.7 and 8.6\,$\mu$m.
\vspace{0.1cm}\\
\end{table*}

\begin{center}
\begin{table*}
  \caption{ List of carbon-rich post-AGB stars in Fig.\ref{pah_6_Galactic}  \label{table-galactic}}
  \begin{tabular}{lllllrrrlllllcc}
\hline
Name & \multicolumn{2}{c}{PAHs} & Sp. or T$_{eff}$ & Ref. \\
 & 6--9\,$\mu$m & 10--14\,$\mu$m \\
\hline
IRAS 01005+7910  & A   & $\beta$ & T$_{eff}$ = 21\,500 K                     &  \citet{Hrivnak:2000dz},   \citet{Klochkova:2002fv}, \cite{Zhang:2010gm} \\
IRAS 04296+3429  & C+D &         & G0I, F3I, T$_{eff}$ = 7\,000 K            & \citet{Bakker:1997wk}, \citet{Klochkova:1999uv}, \\
                                                                             &&&&\cite{VanWinckel:2000uu}, \citet{Volk:2002gd}, \citet{SanchezContreras:2008di}   \\
IRAS 05341+0852  & D   &         & G2Ia, F5I, T$_{eff}$ = 6\,500 K           & \citet{Bakker:1997wk}, \citet{Hrivnak:2000dz}, \cite{VanWinckel:2000uu} \\
                                                                             &&&&\citet{Suarez:2006ba} \\ 
IRAS 07134+1005  & D   &         & F5I, T$_{eff}$ = 7\,250 K                 & \citet{Hrivnak:1989cp}, \citet{Hrivnak:2000dz}, \cite{VanWinckel:2000uu} \\
IRAS 13416-6243  & C   &         & G1I                                       & \citet{Hu:1993td}, \citet{Peeters:2002ci}  \\
IRAS 21282+5050  & A+B & $\delta$+? & Of7-[WC 11]                            & \citet{Cohen:1987iy}   \\
IRAS 22272+5435  & D   & unclassified & G5Ia, , T$_{eff}$ = 6\,500 K         & \citet{1994cmls.conf.....C}, \citet{Hrivnak:2000dz}, \cite{VanWinckel:2000uu}\\
IRAS 22574+6609  & C+D &         & A1--6I                                    & \citet{Hrivnak:2000dz}, \citet{SanchezContreras:2008di}  \\
\hline
\end{tabular} \\
\end{table*}
\end{center}


Figure \ref{pah_lab_fit_spectra} shows the fits to the {\it Spitzer} continuum-subtracted spectra.
Generally, a composite of forty laboratory spectra can reproduce the overall shapes of the observed 10--14\,$\mu$m spectra.

{\it Spitzer} spectra demonstrate the range of spectral variations of PAHs in post-AGB stars, PNe and H{\small II} regions.
The variation of the 10--14\,$\mu$m feature is partly reproduced by composite laboratory PAH spectra.
The fits demonstrated that these features are predominantly attributable to PAHs, and the variations of features (except for class $\delta$) are due to 
the different varieties of PAHs.
We are particularly interested in the fitted results of the class $\gamma$ objects, as the peak wavelength differs from those found in classes $\alpha$ and $\beta$, 
and indeed a combination of the PAH laboratory data can reproduce the 11.4\,$\mu$m peak and the 12 and 13\,$\mu$m broad features.

We found that some observed features were fitted better in one class than in the other classes.
The observed {\it Spitzer} spectra showed a slight shift in the peak wavelength of the 11.2--11.4\,$\mu$m feature (Fig.\,\ref{pah_11_comp}).
The peak appeared at 11.2--11.3\,$\mu$m in class $\alpha$, whereas this was found at 11.4\,$\mu$m in class $\gamma$. 
The composite laboratory spectra tended to reproduce the 11.4\,$\mu$m feature well but not as well as the 11.3\,$\mu$m feature. 
The class $\alpha$ objects were a PN and an H{\small II} region; these have hot central stars, which ionise the surrounding nebulae.
Classes $\beta$ and $\gamma$ were associated with post-AGB stars, i.e. the central stars are cooler and their circumstellar material tends 
to be more neutral (Table\,\ref{table-pagb}).
The poorer fits of the class $\alpha$ objects are probably due to the current limited availability of laboratory data for ionised PAHs.
There were spectra for 30 neutral PAHs in the NASA PAH database, but for only 10 ionised PAHs.
This suggests that a future investment in laboratory measurements of ionised PAHs might improve fits to the class $\alpha$ PAHs.
\citet{Tielens:2008fx} suggested that the typical size of PAHs are 50 carbon atoms, so that laboratory measurements of larger PAHs will
improve the fitting in general.

The class $\delta$ objects showed a broad feature centred at about $\sim$11\,$\mu$m, with a sharp feature at $\sim$11.3\,$\mu$m on top of that.
Our fits show that the broad feature is due to SiC, and it dominated the overall shape, accounting for $>$90\,\% of the energy in the 10--13\,$\mu$m region.
The sharp feature at 11.3\,$\mu$m is due to PAHs, though the fits are not great.
Additionally, though IRAS 05073$-$6752 was classified as $\beta$, its spectrum shows a hint of SiC at a shoulder shortwards of the 11.3\,$\mu$m feature.

\section{Discussions and conclusions}

\begin{figure} 
 \resizebox{\hsize}{!}{\includegraphics*{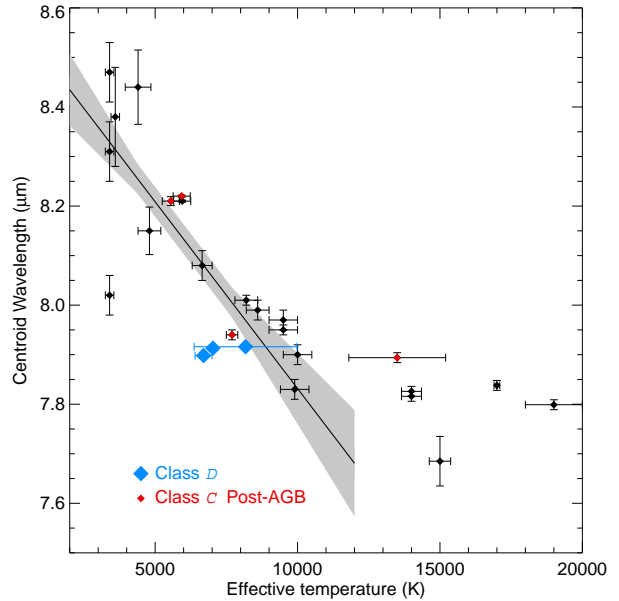}}
  \caption{The central wavelength of $\sim$8\,$\mu$m PAH feature as a function of the temperature.
  Class-$\mathcal{D}$
      \label{pah_cent}}
\end{figure}

{\em Spitzer Space Telescope} spectra of the LMC post-AGB stars
showed a wide variety of features in mid-infrared wavelengths.
The class $\gamma$ type PAH spectra, which have three broad features in the 11--14\,$\mu$m region, and the class-$\mathcal{D}$ spectra, characterised by a 
broad feature peaking at 7.7\,$\mu$m, were found in post-AGB objects in the LMC.  
The PAHs responsible for the $\gamma$ and $\mathcal{D}$ type spectra tend to be found more in post-AGB stars than in any other types of objects that carry PAHs.

Previous studies of PAHs in Galactic objects 
have found that the central wavelength of $\sim$8\,$\mu$m PAH feature correlates
with the effective temperature of the central star
\citep{Sloan:2007hf, Boersma:2008kr, Smolders:2010p29274}.
Objects with our newly found class-$\mathcal{D}$ PAHs appear to be outliers from this correlation (Fig.\ref{pah_cent}).
In this figure, grey shadow shows the linear fit to the correlation  \citep{Smolders:2010p29274},
and the spectral types of class-$\mathcal{D}$ objects are converted to the effective temperatures \citep{2000asqu.book.....C}.
The previously reported correlation predicts that the stars with effective temperature of 6100-9730K should show
a peak of PAHs at about 8.2\,$\mu$m.
However, these class-$\mathcal{D}$ objects have the peak at 7.8--7.9\,$\mu$m, much shorter than the  expected wavelength from the effective temperature.
This correlation has been thought to be caused by a change of PAH compositions, as the central stars evolve and their increasing energetic radiation affects the PAHs.
Our finding of class-$\mathcal{D}$ objects suggest that the compositions of PAHs might be more variable than previously thought.

A wide variety of PAH features have been found in our study, but mainly in the post-AGB stars.
By the time the central star evolves into the PN phase, PAHs may have been processed and the profiles changed into class A or B at 6--9\,$\mu$m and 
class $\alpha$ at 10--14\,$\mu$m.
PAHs in Post-AGB stars and PNe seem to have been altered by UV radiation as the star evolves towards higher surface temperature following the AGB stage, and 
PAH features change their shape.
By the time PAHs are ejected from carbon-rich AGB stars into the ISM, PAHs would already be processed, and spectral characteristics would not differ so much from 
what is found in the ISM.

The question remains why low metallicity impacts on the profiles of PAHs emitted from the ISM of galaxies \citep{Draine:2007de},
although little difference has been found in PAHs emitted from PNe and post-AGB stars hosted by low-metallicity galaxies.
There might be a slight difference, as no class $\mathcal{C}$ PAHs were found in LMC
post-AGB stars, but confirmation is required with a larger sample in future.
One possibility is that UV radiation in the general ISM and PNe might change the profiles of PAHs \citep{Lebouteiller:2011cx}.
An alternative process might be that PAHs in the ISM could be further re-processed within the ISM, changing the composition \citep{Matsuura:2013js}.
This contrast of the ISM and PN PAH profiles could have potential in resolving
how dust could be excited and processed in the circumstellar nebulae and ISM.

\section{acknowledgements}
M.M. thanks Prof. I. D. Howarth for his inputs about the classifications of high-mass stars.
We thank Dr. J.P. Seale for providing the Herschel measurements of IRAS 05189$-$7008 fluxes.
Support for this research was provided by NASA through contract 1378453 issued by JPL/Caltech.
This publication makes use of data products from the Two Micron All Sky Survey, which is a joint project of the University of Massachusetts and the Infrared Processing 
and Analysis Center/California Institute of Technology, funded by the National Aeronautics and Space Administration and the National Science Foundation. This research 
has made use of the SIMBAD database, operated at CDS, Strasbourg, France.
R.Sz. acknowledges support from Marie Curie Actions IRSES (project No. 269193) of EU and from Polish NCN grant 2011/01/B/ST9/02031.
FK acknowledges financial support from the National Science Council under grant number NSC100-2112-M-001-023-MY3.
R.S.'s contribution to the research described here was carried out at the Jet Propulsion
Laboratory, California Institute of Technology, under a contract with NASA, and supported
via an award issued by JPL/Caltech in support of his Spitzer Guest Observer program.


\bibliography{pagb}

\bibliographystyle{mn2e}

\appendix

\section{Colour-colour and colour-magnitude diagrams} \label{appendix-cc}

We have already presented one of the representative colour-colour diagrams used for the target selection in Sect.\,\ref{cc-diagrams}; 
the remaining diagrams are presented here.

\subsection{Colour-colour diagram $J-$[8.0] vs [8.0]$-$[24]}

\begin{figure}
  \rotatebox{90}{ \begin{minipage} {6.5cm} 
   \resizebox{\hsize}{!}{\includegraphics*[70, 131][497, 673]{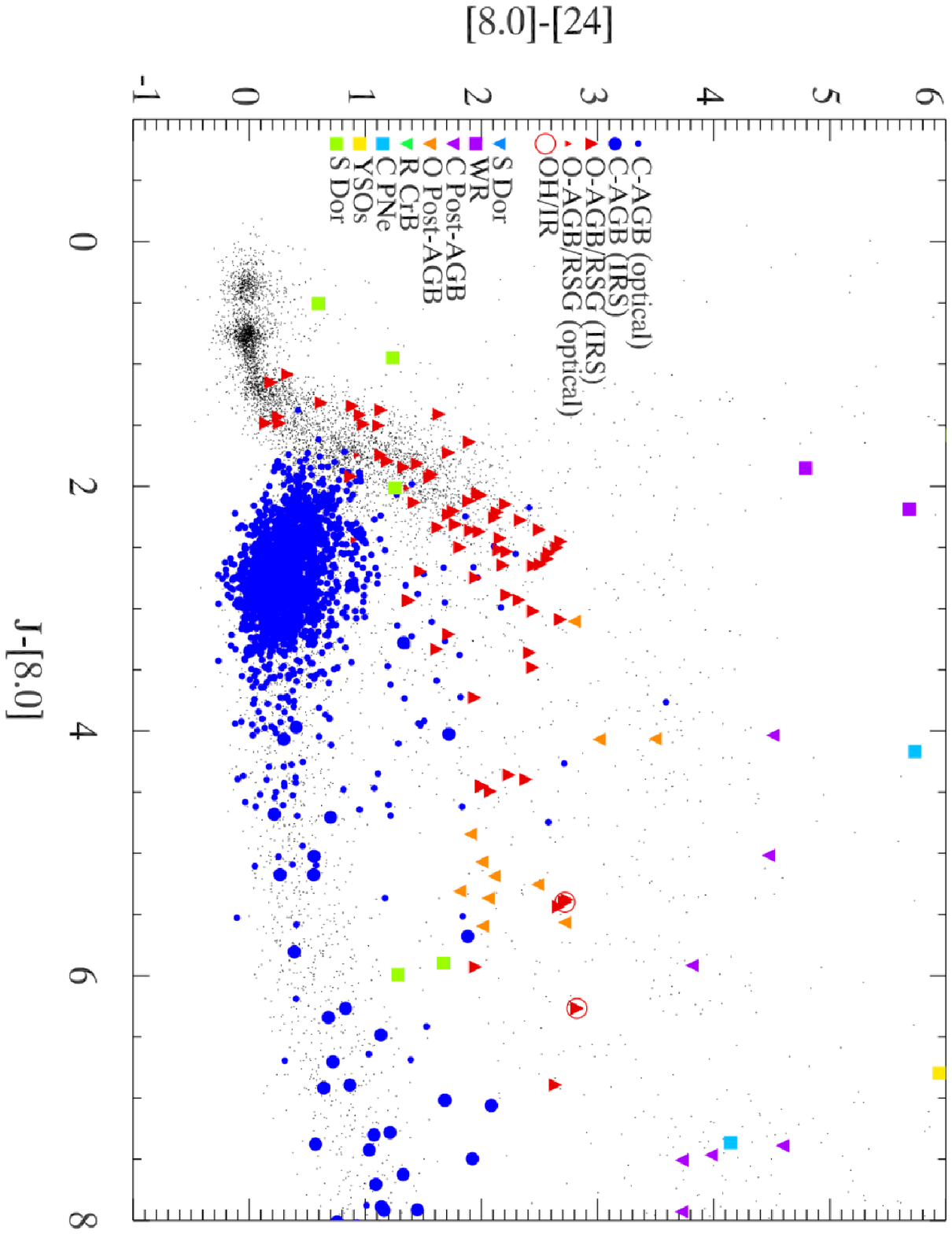}}  \end{minipage}}
   \caption{ The $J-[8.0]$ vs $[8.0]-[24]$ colour-colour diagram.
   Post-AGB stars fall in the red part of the diagram ($J-[8.0]>3.0$ and $[8.0]-[24]>1.8$).
   \label{cc_j8_824}}
\end{figure}

Figure\,\ref{cc_j8_824} shows the $J-$[8.0] vs [8.0]$-$[24] colour-colour diagram.
Post-AGB stars have red mid-infrared colours, $[8.0]-[24] > 1.8$.

\citet{2011A&A...530A..90V} used the $[8.0]-[24]$  colour to find post-AGB candidates, and set the criteria of post-AGB to be $[8.0]-[24] > 1.384 $.
Figure\,\ref{cc_j8_824} shows that all of our post-AGB stars fulfil this condition, but our sample proposes a redder cutoff at $[8.0]-[24] > 1.8$.

Although the size of the current sample is limited, Fig.\,\ref{cc_j8_824} suggests that the $[8.0]-[24]$ colour can distinguish three types of objects; 
carbon-rich post-AGB stars, oxygen-rich post-AGB stars, and carbon-rich AGB stars.
Oxygen-rich post-AGB stars have a colour range of  $1.8 < [8.0]-[24] < 3.5 $, whereas carbon-rich post-AGB stars have $[8.0]-[24] > 3.5 $, and carbon-rich AGB stars 
have bluer colours with $[8.0]-[24]< 2$.
The difference is due to the dust properties of these three types of objects.

We used the criteria of $J-[8.0]>5$ and $[8.0]-[24]>0.5$ to select carbon-rich AGB stars.
We found that the colour distributions of carbon-rich post-AGB stars are slightly bluer in $J-[8.0]$ ($>4$), but redder in $[8.0]-[24]$ ($>3.5$).

 \subsection{Colour-magnitude diagrams}

\begin{figure}
  \rotatebox{90}{ \begin{minipage} {6.5cm} 
  \resizebox{\hsize}{!}{\includegraphics*[75, 124][500, 690]{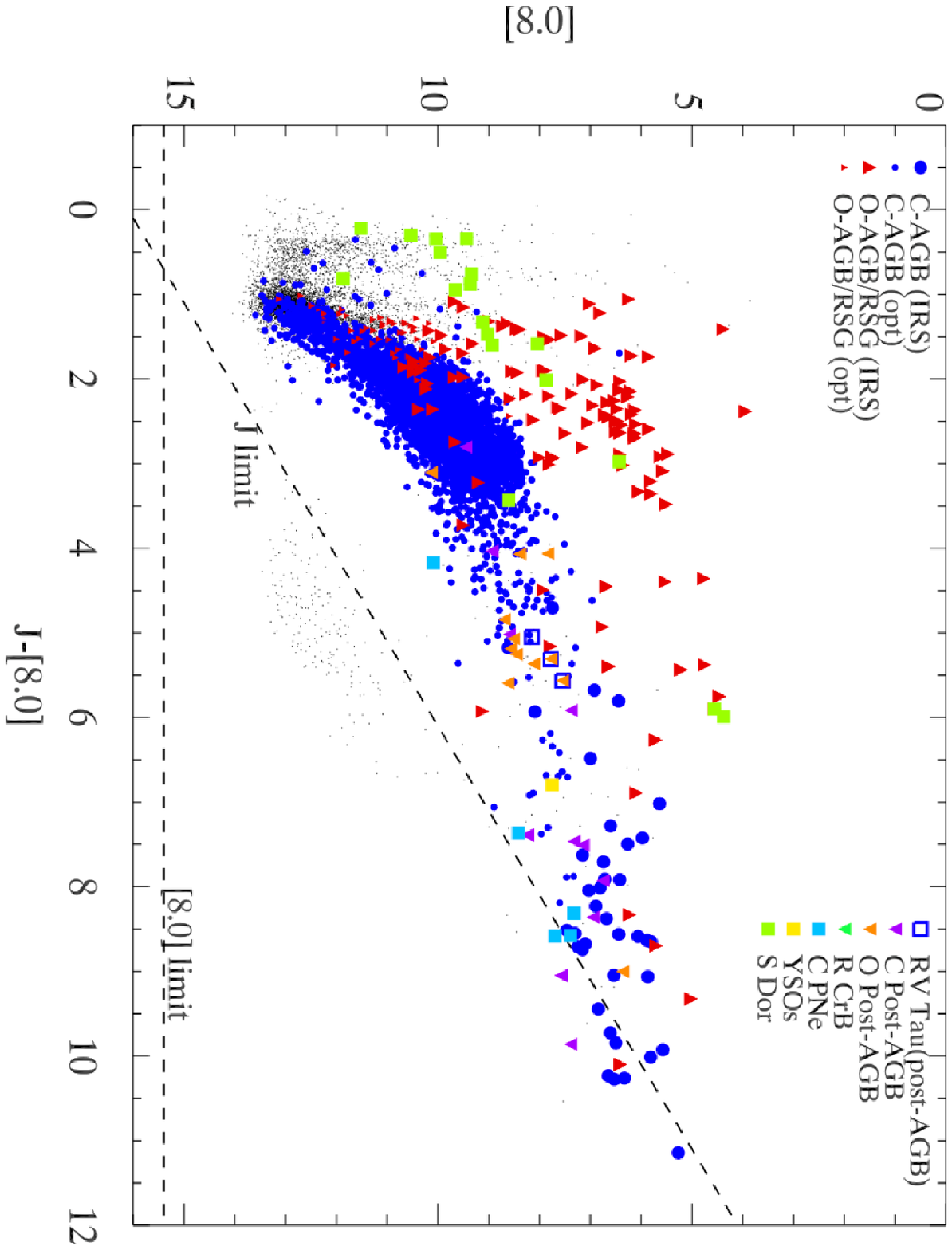}}  \end{minipage}} 
   \caption{Colour-magnitude diagram $J-[8.0]$ vs $[8.0]$, which was used for part of the target selection.
   The oxygen-rich post-AGB stars fall in a small range, $3<J-[8.0]<6$ and $7<[8.0]<10$, with the exception of one star at $J-[8.0]\sim8.5$. 
Carbon-rich post-AGB stars are found in the range of $J-[8.0]>7$ and $6<[8.0]<9$.
   Post-AGB stars tend to have similar colours to red AGB stars in this diagram.
   {\it Spitzer} and 2MASS detection limits are plotted as dashed lines.
   \label{magJ8_8}}
\end{figure}

\begin{figure}
  \rotatebox{90}{ \begin{minipage} {6.5cm} 
  \resizebox{\hsize}{!}{\includegraphics*[75, 124][500, 690]{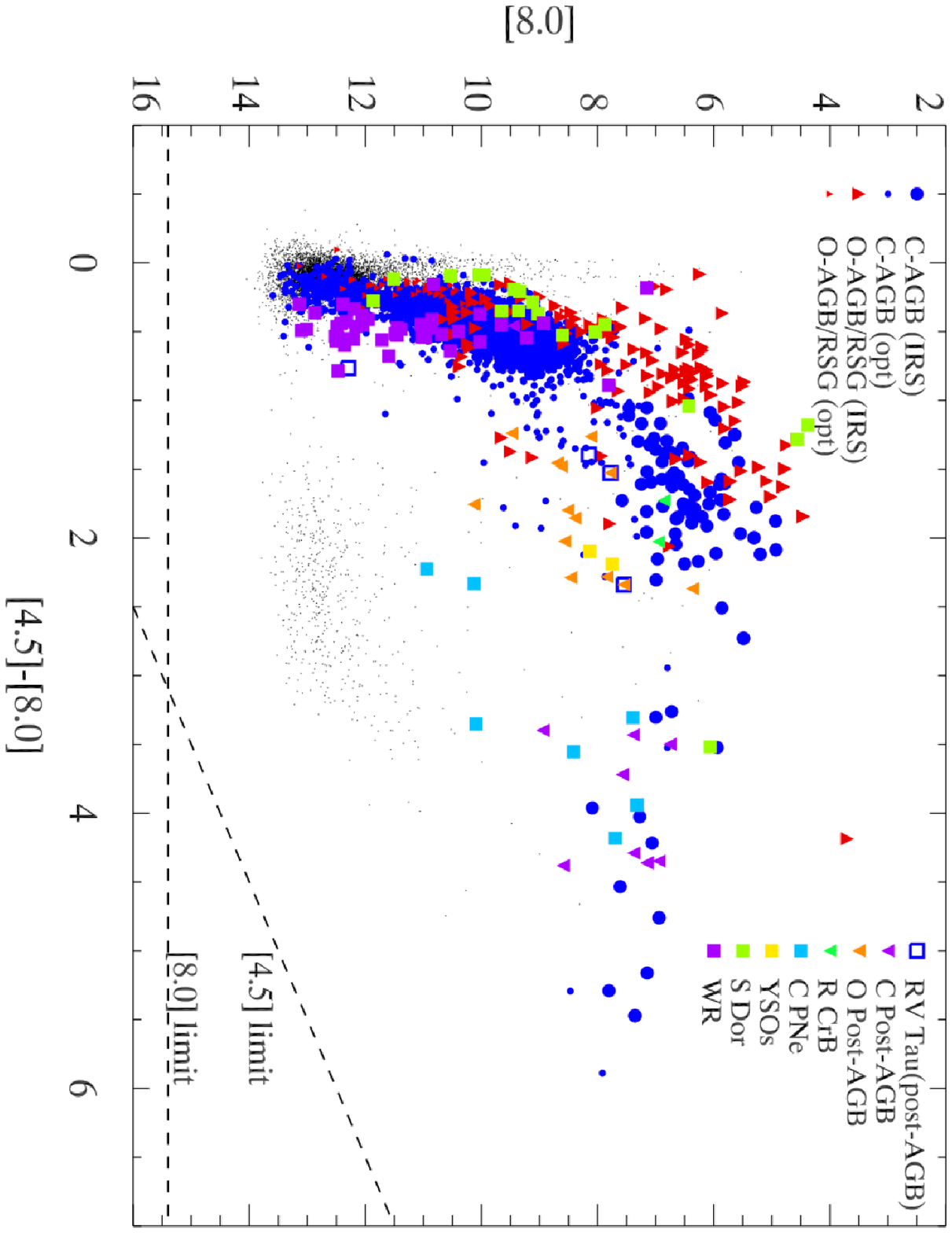}}  \end{minipage}} 
   \caption{Colour-Magnitude diagram $[4.5]-[8.0]$ vs $[8.0]$, which was used for part of the target selection.
     Oxygen-rich post-AGB stars fall in the ranges $1<[4.5]-[8.0]<2.5$ and $6<[8.0]<10$, fainter than infrared-selected oxygen-rich AGB stars and supergiants.
    Carbon-rich post-AGB stars have much redder colours than oxygen-rich post-AGB stars, having $3<[4.5]-[8.0]<5$. Carbon-rich post-AGB stars and carbon-rich PNe have 
similar colours.
   \label{mag_48_8}}
\end{figure}

We examine the colour-magnitude diagrams, used for the selection of post-AGB stars.
Fig.\,\ref{magJ8_8} shows the $J-[8.0]$ vs [8.0] colour-magnitude diagram: 
post-AGB stars have $J-[8.0]>4$, except for one object, and $5<[8.0]<10$.
Similarly, Fig.\,\ref{mag_48_8} shows the $[4.5]-[8.0]$ vs [8.0] colour-magnitude diagram, where carbon-rich post-AGB stars are found at $[4.5]-[8.0]>3$.
However, these diagrams suffer from contaminations by other types of object, particularly carbon-rich AGB stars.
After the selection of post-AGB candidates, spectroscopic follow-up observations are essential  reveal their true nature.

\subsection{The $[3.6]-[4.5]$ vs $[4.5]$ colour-magnitude diagram}

\begin{figure}
  \rotatebox{90}{ \begin{minipage} {6.5cm} 
  \resizebox{\hsize}{!}{\includegraphics*[75, 124][500, 690]{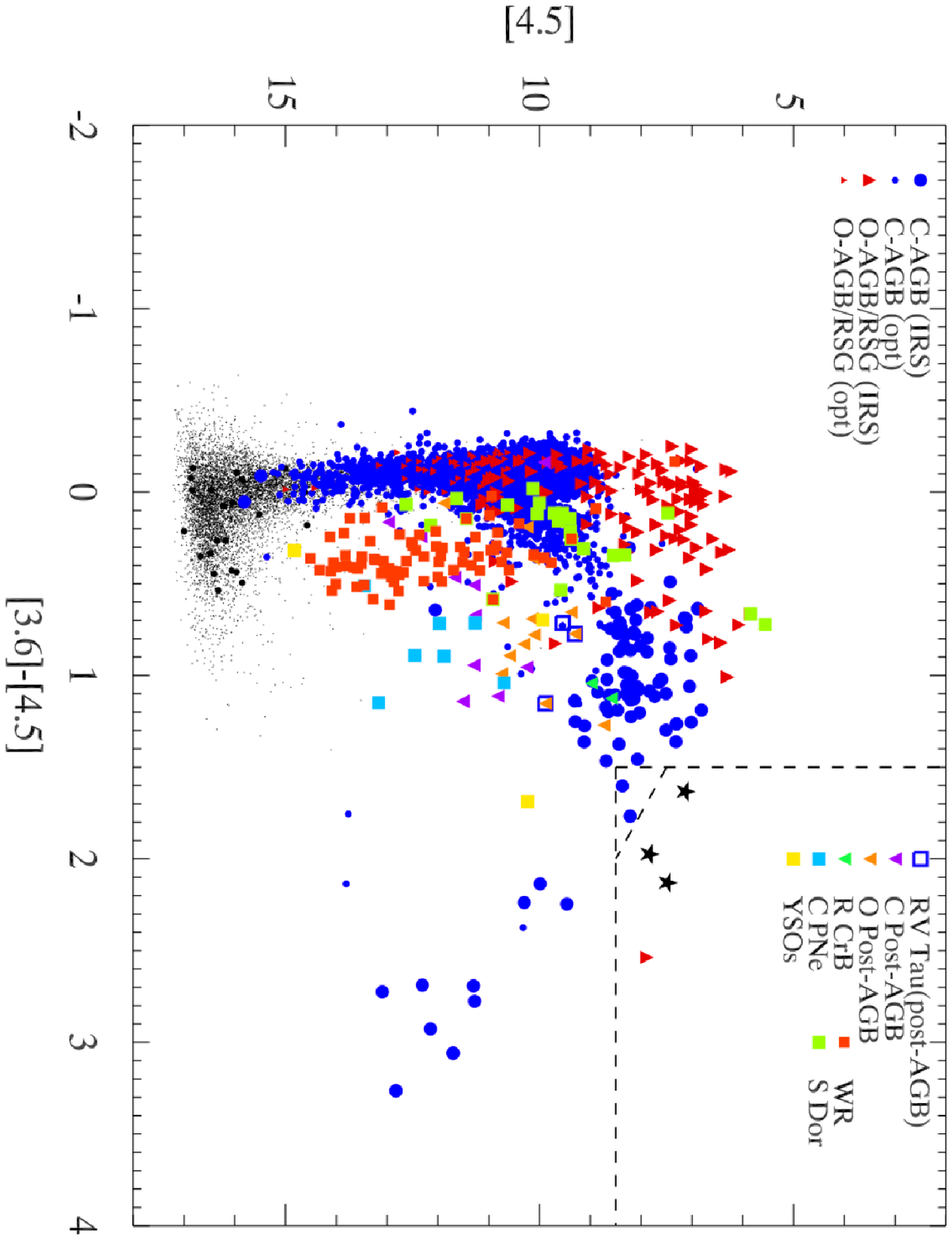}}  \end{minipage}} 
   \caption{Colour-Magnitude diagram $[3.6]-[4.5]$ vs $[4.5]$, which has been used for object classifications.
   The box shows the region where the progenitors of SN2008S can be found. Among several possibilities, this diagram suggests that 
carbon-rich AGB stars or oxygen-rich RSGs are the most likely object types.
   \label{mag_34_4}}
\end{figure}

The colour-magnitude diagram $[3.6]-[4.5]$ vs $[4.5]$ is often used to identify AGB stars/post-AGB stars in nearby galaxies
\citep[e.g.][]{Thompson:2009cb},
because these two {\it Spitzer} wavebands tend to contain more stars than the other two {\it Spitzer} IRAC ([5.8] and [8.0]) bands.

The particular interest of the  $[3.6]-[4.5]$ vs $[4.5]$ diagram is because it was used to diagnose the evolutionary stage of the progenitor star of 
supernova (SN) 2008S, which could be an AGB star or a post-AGB star \citep{Wesson:2009dk, Prieto:2009hy, Smith:2009co, Botticella:2009jy}.
The explosion of SN 2008S was detected in the spiral galaxy NGC 6946, and its progenitor star was found in the Spitzer image of NGC 6946 taken before the explosion.
From Spitzer photometric data, \citet{Thompson:2009cb} proposed that the progenitor star should be a high-mass star but within the relatively low mass range 
(8--12\,$M_{\sun}$).
\citet{Wesson:2009dk} suggested that it could be a carbon-rich AGB star or super-AGB star with a mass range of 6--10\,$M_{\sun}$.
\citet{Prieto:2008bv} and \citet{Prieto:2009hy} proposed that the progenitor of SN 2008S is a carbon-rich post-AGB star, from the presence of Ca{\small II} emission in the 
optical spectra and from the infrared colour showing a carbon-rich circumstellar envelope.

The carbon-rich AGB/post-AGB possibility for the progenitor of SN 2008S can be assessed by comparing the colour magnitude diagrams of our 
spectroscopically classified objects in the LMC.
The progenitor of SN 2008S had a colour of $[3.6]-[4.5]>1.5$ and the absolute luminosity of $M_{4.5}<-10$ \citep{Prieto:2008bv}.
This 4.5\,$\mu$m absolute magnitude corresponds to an apparent magnitude of $[4.5]<8.5$ for the LMC distance, assuming the distance module of 18.5\,mag for the LMC.
The corresponding colour and magnitude of the SN2008S progenitor scaled to the LMC distance is indicated as the dashed-line box in Fig.\,\ref{mag_34_4}.
Two carbon-rich stars and one oxygen-rich red supergiant,  which is \citep[IRAS 05346$-$6949; ][]{Elias:1986fo}, are found within this box.
No known S\,Dor or WR stars fall within this box, neither are there any post-AGB stars or PNe.
There are three unclassified objects, indicated by `star' symbols, within the box.
From this colour-magnitude diagram, the colour corresponding to the progenitor of SN 2008S, seems to have been either a carbon-rich AGB star or an oxygen-rich red 
supergiant, but is less likely to have been a post-AGB star.

\section{Demonstration of the small influence of the continuum on the peak wavelengths of the 6--9\,$\mu$m features} \label{appendix-cont}

\begin{figure}
  \resizebox{\hsize}{!}{\includegraphics*{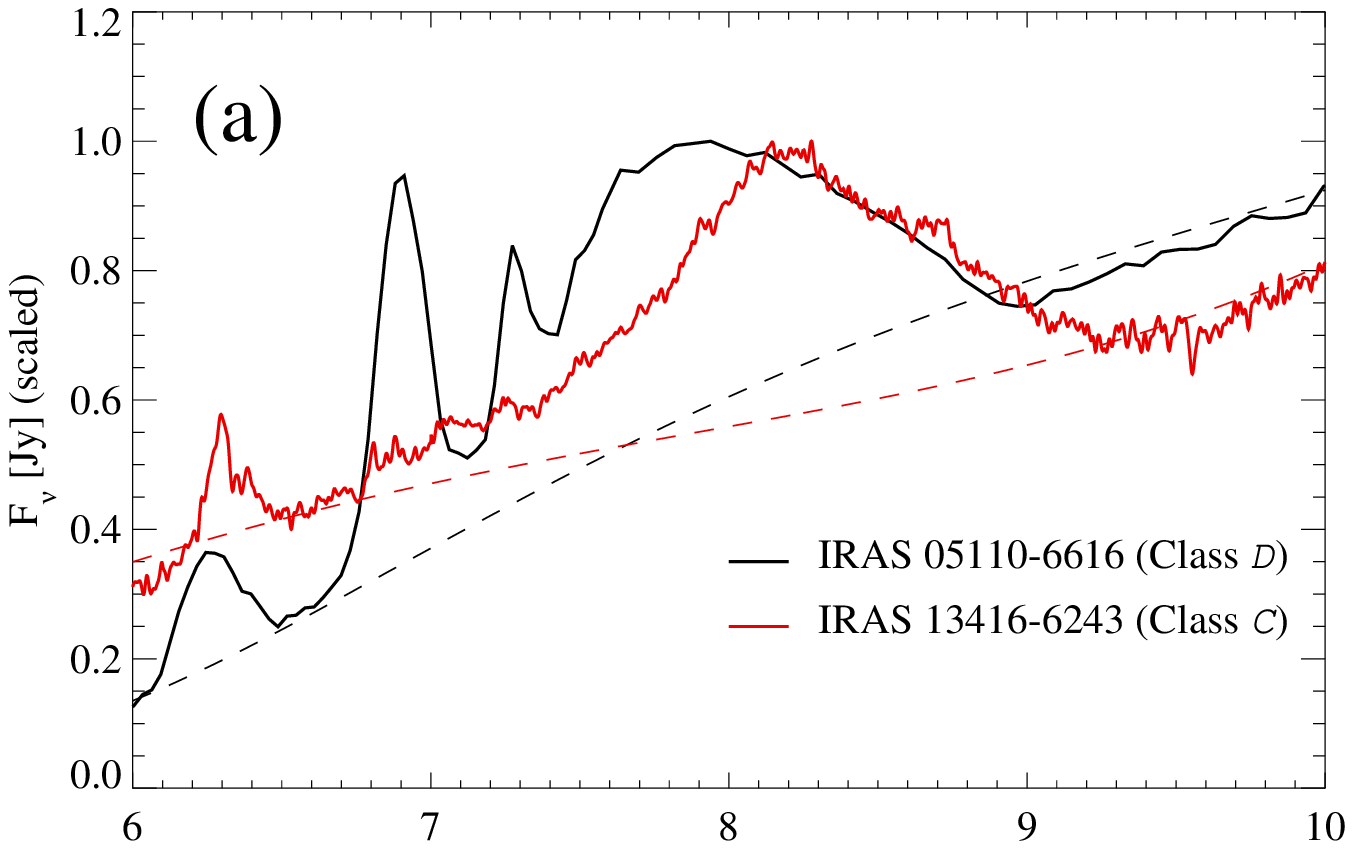}} 
  \resizebox{\hsize}{!}{\includegraphics*{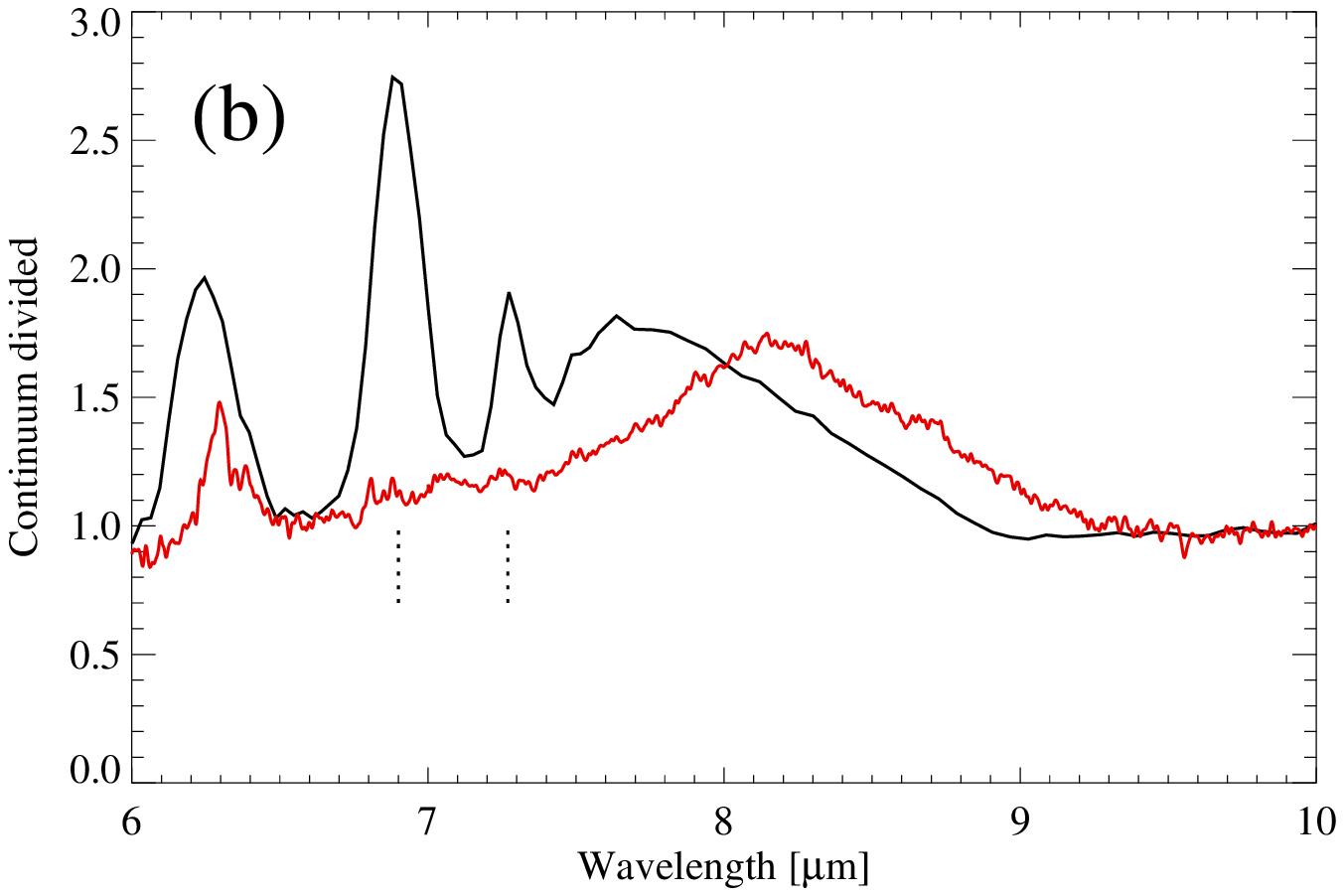}} 
   \caption{Demonstration of the small influence of the continuum definitions on the peak wavelengths of the broad 7.5--9\,$\mu$m features.
   Panel (a) shows the observed spectra of IRAS 05110$-$6616 (LMC class-$\mathcal{D}$ object) and IRAS 13416$-$6243 (Galactic class-$\mathcal{C}$ object), along 
with their continua in dashed lines. 
The fluxes are scaled to the peak value within the 6--9\,$\mu$m range.
The spectrum of  IRAS 05110$-$6616 indicates the peak at about 7.8--7.9\,$\mu$m, whereas the peak is found at 8.2\,$\mu$m for IRAS 13416$-$6243. 
   Panel (b) indicates the continuum-divided spectra, instead of continuum-subtracted spectra, which were used in the PAH analysis.
   Comparisons of these two spectra show the difference in the peak wavelengths of these two spectra remain even in continuum-divided spectra.
Note that the spectrum of IRAS\,05110$-$6616 has in addition two narrow features at $\sim$6.9 and $\sim$7.3\,$\mu$m, apart from the broad feature 
at 7.5--9\,$\mu$m region.
    \label{spec_cont2}}
\end{figure}

The continuum subtraction may potentially distort the features, raising concern that the PAH classification of the profiles $\mathcal{C}$ and $\mathcal{D}$ of the 
broad 7.5--9\,$\mu$m features might be caused by poor continuum subtraction.
In order to establish that the distinction of profiles $\mathcal{C}$ and $\mathcal{D}$ is indeed present, Figure \ref{spec_cont2} shows the observed spectra and the 
continuum divided spectra of the representative class $\mathcal{C}$ and class $\mathcal{D}$ objects.
IRAS 13416$-$6243 is a Galactic post-AGB star with a class $\mathcal{C}$ PAH profile in the 6--9\,$\mu$m range, showing a broad single peak at 8.22\,$\mu$m. 
IRAS 05110$-$6616 is an LMC class $\mathcal{D}$ object, that also has a broad feature between 7.5--9\,$\mu$m range, but with the peak at a shorter wavelength 
than in class $\mathcal{C}$.

In our analysis, we used the continuum-subtracted spectra. 
In order to demonstrate that the continuum subtraction did not create an artificial distinction between class $\mathcal{C}$ and class $\mathcal{D}$ profiles, 
Figure\,\ref{spec_cont2} (b) displays the continuum divided spectra. 
A rising continuum, if the continuum is more dominant than the features,
could potentially shift the peak wavelengths of the features towards longer wavelengths, but this effect could be cancelled out better in continuum-divided spectra 
than in continuum-subtracted spectra.
Figure \ref{spec_cont2} (b) demonstrates that class $\mathcal{C}$ and class $\mathcal{D}$ objects clearly have different peak wavelengths
of the 7.5--9\,$\mu$m broad feature.

\section{Optical spectra} \label{appendix-optical}

\begin{figure}
\rotatebox{270}{ \begin{minipage} {11.5cm}
   \resizebox{\hsize}{!}{\includegraphics*[48, 198][536,560]{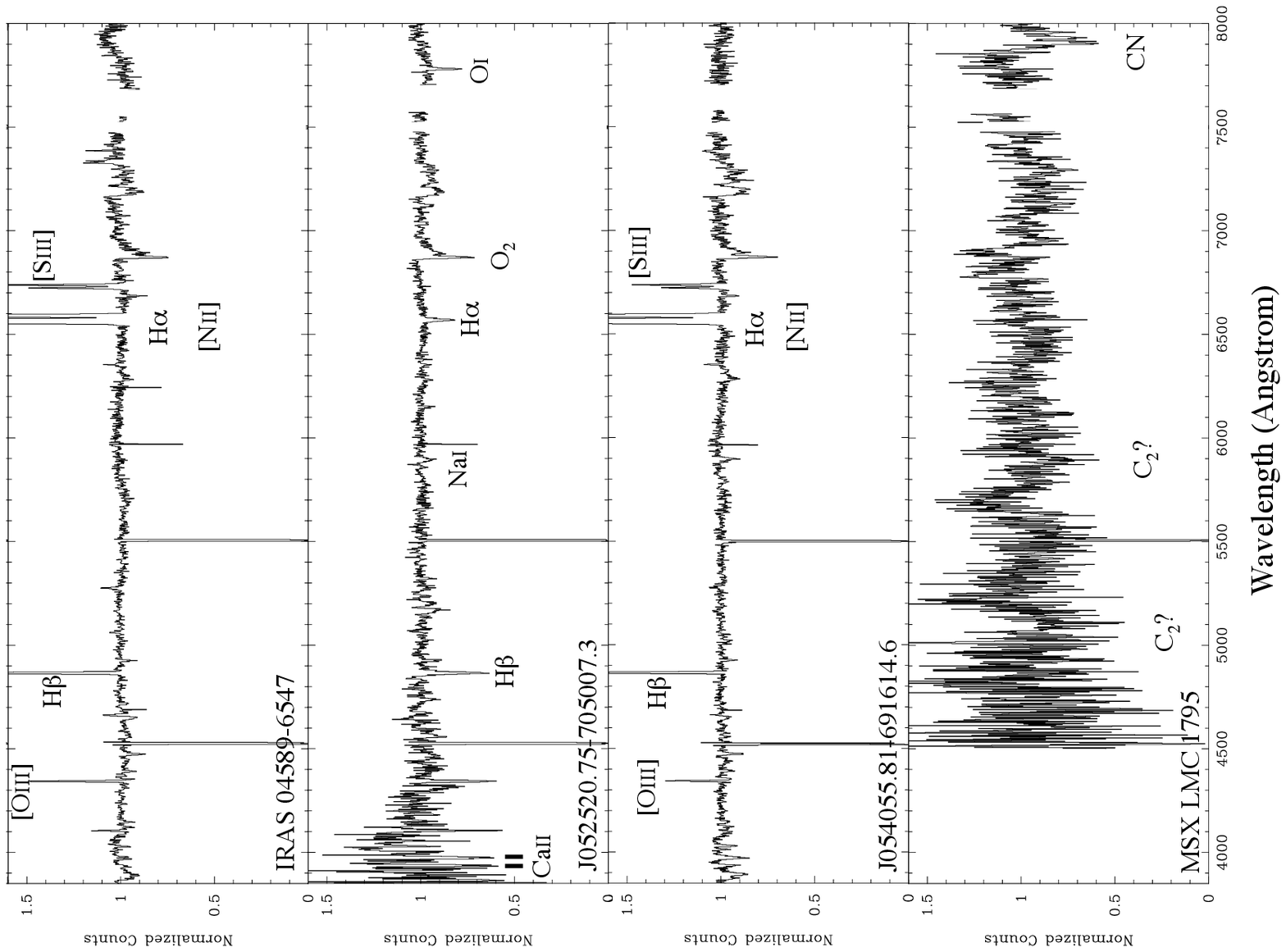}} 
\end{minipage}}
   \caption{Normalised optical spectra of four objects, IRAS 04589$-$6547, J052520.75$-$705007.3, J054055.81$-$691614.6 and MSX\,LMC\,1795.
       \label{spec_optical}}
\end{figure}

Optical spectra of four objects (IRAS 04589$-$6547, J052520.75$-$705007.3, J054055.81$-$691614.6 and MSX\,LMC\,1795) were obtained with 
the Blanco 4-metre telescope using the RC Spectrograph on 27 Jan 2010,
and their spectra are shown in Figure\,\ref{spec_optical}. 
The spectrograph employed the KPGL2-1 grating (316 lines~mm$^{-1}$), and 
the resulting spectrum was imaged by the blue Air Schmidt Camera and Loral 3k (\# 1) CCD.  
The 150~$\mu$m (1\arcsec) wide long-slit was oriented in the east-west direction.  
The resulting spectra were sampled with $\sim$2\AA~pixel$^{-1}$, 
and have an effective spectral resolution of $\sim$750 ($\sim$8.5 \AA) at H$\alpha$ (as measured from unresolved sky lines).
The observations were reduced using the ccdproc and twodspec packages in IRAF.  
Each spectrogram was overscan and bias corrected, then the wavelength calibration, based on exposures of a HeNeAr lamp taken just prior to or after those for the 
source, and a spatial distortion correction were applied to each spectrogram.

The spectrum of J052520.75$-$705007.3 in Figure\,\ref{spec_optical}
shows strong Ca{\small II} H and K, weak G band, sharp weak H line absorption and strong O{\small I} 7774$\AA$, 
giving a spectral type of F2--F5I, with reference to the spectral atlas of 
\citet{Jacoby:1984ep}. This is significantly later than the type A1Ia \citep{2011A&A...530A..90V}, based on a red spectrum with an undetectable CaII triplet.

The spectrum of MSX\,LMC\,1795 has a moderately strong red CN band and a possible C$_2$ Swan band at 5165\,$\AA$ and CN,
showing it is a carbon star, consistent with our classification of the mid-infrared spectrum as typical of an R\,CrB star. The lack of coverage in the blue spectral 
region and the low S/N preclude inspection of the features diagnostic of an R\,CrB star. 
The 2MASS, 6X2MASS and IRSF photometry show a considerable spread, because of a faint blue companion at about 1\,arcsec distance which affects $J$ as well as 
intrinsic variability. All three datasets show it is a very red star which clearly lies in the R\,CrB rather than the DY\,Per zone in the two colour diagram 
\citep{Tisserand:2009kba}, so it is an R\,CrB star and not a DY\,Per star or a C-rich AGB star.

The optical spectra of IRAS 04589$-$6547 in our Figure\,\ref{spec_optical} and in \citet{2011A&A...530A..90V} give a spectral type 
O9ep/B1-2ep, while that of J054055.81$-$691614.6 is O--B. Both stars have intense H$\alpha$ emission, and the equivalent widths of the emission lines have been 
measured with one or more of the instruments described in this section and the Unit Spectrograph on the 1.9m Radcliffe reflector at SAAO, which gave a resolution of 
4-5\,\AA\,depending on wavelength. The EW values are given for IRAS 04589$-$6547 first: [OII] 3727\,\AA: 5.4, 0.8, H$\delta$: 1.0, 0.3, H$\gamma$: 3.4, 1.9, 
H$\beta$: 15.2, 9.4, H$\alpha$: 132, 102, [NII] 6548\,\AA: 20, 24, [NII] 6584\,\AA: 62, 76, [SII] 6717\,\AA: 3.1, 2.0, [SII] 6731\,\AA: 5.1, 3.3, 
OI 8446\,\AA: 1.4, 1.8, [SIII] 9069\,\AA: 23, 10. Observations on different dates give no indication of variability in these line strengths. 

\begin{figure}
   \resizebox{\hsize}{!}{\includegraphics*[41,168][571,712]{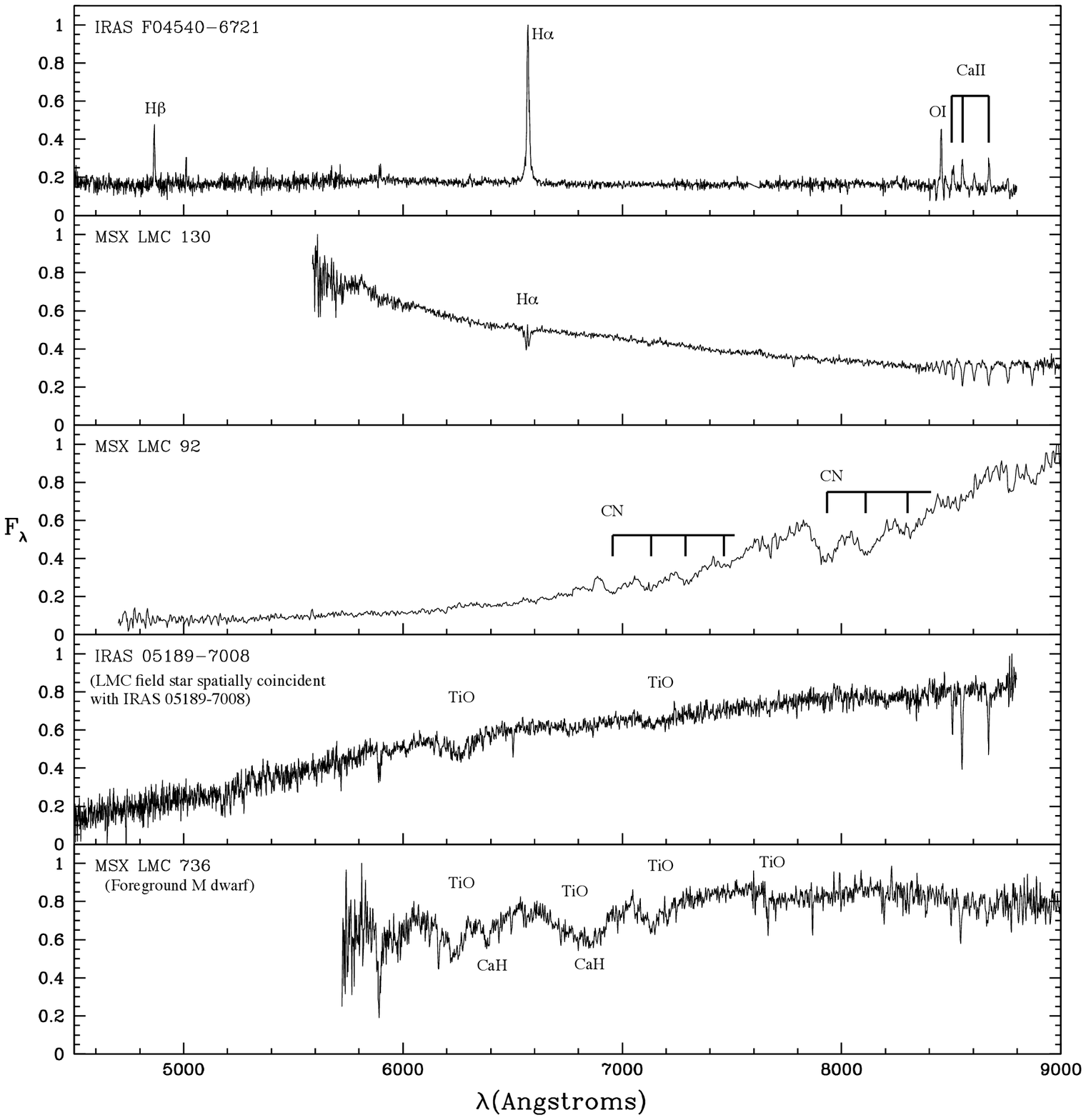}} 
   \caption{Optical spectra for 5 objects, showing relative $F_{\lambda}$ versus
$\lambda$. The optical spectra of IRAS 05189-7008 and MSX\,LMC\,736 are not those
 of the mid-infrared sources, but correspond to field stars in the same line of sight.
       \label{spec_optical2}}
\end{figure}

Optical spectra of 5 objects were observed with the Anglo-Australian
Telescope (AAT) or the 2.3m telescope of the Australian National
University at Siding Spring Observatory, and their spectra are shown in Fig.\,\ref{spec_optical2}.  
The spectra of IRAS 05189$-$7008 and IRAS F04540$-$6721 were taken with the AAT
on 28 and 29 Oct 2010, respectively, using the multi-fibre
AAOmega spectrograph \citep{Smith:2004di} with a resolution of
$\sim$1300.  The spectra of MSX\,LMC\,130 and MSX\,LMC\,736 were taken on
13 Jan 2000 using the Dual-Beam Spectrograph (DBS)
\citep{Rodgers:1988dv} on the 2.3m telescope.  
The spectra have a resolution of $\sim$2000.  
MSX\,LMC\,92 was observed with the same
telescope and instrument on 25 Dec 2003 at a lower resolution of
$\sim$1000.  In all the spectra, telluric features have been removed
by dividing the spectra by the spectrum of a featureless white dwarf.

IRAS F04540$-$6721 is an emission line star showing emission lines of H{\small I}
(H$\alpha$ and H$\beta$), O{\small I} (8446\,\AA), Na{\small I} (5889,5895\,\AA) and
Ca{\small II} (8498, 8542, 8662\,\AA). Possible [O{\small III}] (5007\,\AA) could be from the LMC field. 
MSX\,LMC\,130 has a spectral type of about A7, with H$\alpha$ in emission in the core of the absorption line.
MSX\,LMC\,92 is a cool carbon star as shown by the characteristic CN bands.

The optical spectra of two stars were found to be those of field stars which are spatially coincident with the mid-infrared sources.
The optical counterpart to IRAS 05189$-$7008 shows weak TiO bands near 6250\,\AA\ and
7050\,\AA, so it is an O-rich star of type near K4.  $JHK$$_{s}$ photometry from the IRSF, 2M and 6X2M give concordant 
results, $J$--$H$ = 0.84, $H$--$K$$_{s}$ = 0.19, in agreement with the type. The offset from the catalogue position of 1.0\,arcsec may be significant.
The optical counterpart to MSX\,LMC\,736 shows strong CaH bands near 6380 and 6850\,\AA, as well as TiO
bands, and it appears to be an M dwarf (M1V), located in the foreground of the LMC. The positional offset is zero in this case, and the IRSF, 2M and 6X2M 
photometry obtained at different epochs give ranges of 0.01, 0.08 and 0.53\,mag at $J$, $H$ and $K_{s}$ respectively. This is consistent with an increasing 
contribution from a variable AGB star towards longer wavelength. The near infrared photometry of these two stars was not used in Fig. 1, in view of the 
contamination by field stars.

\label{lastpage}

\end{document}